\def\astroph{1}
  \ifnum\astroph=0
     \documentclass[manuscript]{aastex}          
  \else
    \documentclass{emulateapj}
   \fi

  \usepackage{epsfig,amsmath,natbib}
  \usepackage[dvips]{color}
  \usepackage[pdftex]{hyperref}
  \usepackage{verbatim}
  
  \newcommand{\F}{F}
  \newcommand{\Flam}{\F(\lambda)}
  \newcommand{\T}{\mathcal{T}}
  \newcommand{\nhi}{n_{\textrm{{\scriptsize H}{\tiny \hspace{.1mm}I}}}}
  \newcommand{\Msun}{M_\odot}

  \newcommand{\Sec}[1]{Section \ref{sec:#1}}
  \newcommand{\Fig}[1]{Figure \ref{fig:#1}}
  \newcommand{\Tab}[1]{Table \ref{tab:#1}}
  \newcommand{\Eq} [1]{Equation (\ref{eq:#1})}
  \newcommand{\zre}{z_{\mathrm{re}}}

  \hypersetup{colorlinks,%
              citecolor=black,%
              filecolor=black,%
              linkcolor=black,%
              urlcolor=blue,%
              pdftex}

\begin{document}
 \slugcomment{2nd draft}
\shorttitle{Intergalactic Transmission and its Impact on the Ly$\alpha$ Line}
\shortauthors{Laursen et al.}
\title{Intergalactic Transmission and its Impact on the Ly$\alpha$ Line}
\author{Peter Laursen\altaffilmark{1,2},
        Jesper Sommer-Larsen\altaffilmark{1,3,4}
        and Alexei O. Razoumov\altaffilmark{5}}
\altaffiltext{1}{Dark Cosmology Centre, Niels Bohr Institute, University of
                 Copenhagen, Juliane Maries Vej~30, DK-2100, Copenhagen {\O},
                 Denmark; email: pela@dark-cosmology.dk,
                 j.sommerlarsen@gmail.com.}
\altaffiltext{2}{Oskar Klein Centre, Dept.~of Astronomy, Stockholm University,
                 AlbaNova, SE-10691 Stockholm, Sweden.}
\altaffiltext{3}
{Excellence Cluster Universe, Technische Universit\"at M\"unchen,
Boltzmannstra{\ss}e 2, 85748 Garching, Germany.}
\altaffiltext{4}
{Marie Kruses Skole, Stavnsholtvej 29-31, DK-3520 Farum, Denmark.}
\altaffiltext{5}{SHARCNET/UOIT,
2000 Simcoe Street N.,
Oshawa ON L1H7K4, Canada;
email: razoumov@sharcnet.ca.}

\begin{abstract}
We study the intergalactic transmission of radiation in the vicinity of
the Ly$\alpha$ wavelength. Simulating sightlines through the intergalactic
medium (IGM) in detailed cosmological hydrosimulations, the impact of the IGM
on the shape of the line profile from Ly$\alpha$ emitting
galaxies at redshifts 2.5 to 6.5 is investigated.
In particular we show that taking into account the
correlation of the density and velocity fields of the IGM with the
galaxies,
the blue part of the spectrum may be appreciably
reduced, even at relatively low redshifts. This may in some cases provide
an alternative to the often-invoked outflow scenario, although it is
concluded that this model is still a plausible explanation of the many
asymmetric Ly$\alpha$ profiles observed.

Applying the calculated wavelength dependent transmission to simulated spectra
from Ly$\alpha$ emitting galaxies, we derive the fraction of photons that are
lost in the IGM, in addition to what is absorbed internally in
the galaxies due to dust.

Moreover, by comparing the calculated transmission of radiation blueward of the
Ly$\alpha$ line
with corresponding observations, we are able to constrain the epoch when the
Universe was reionized to $z \lesssim 8.5$.
\ \\

\end{abstract}


\section{Introduction}

The past decade has seen a rapid increase in the use of Ly$\alpha$ as a
cosmological tool. A plethora of physical characteristics of young galaxies
and the intergalactic medium (IGM) that separate them manifest themselves in
the shape, the strength, and the spatial distribution of both Ly$\alpha$
emission and absorption lines. 

Since generally the evolution of astrophysical objects happen on timescales
much longer than a human lifetime, numerical simulations are particularly
well-suited for exploring the Universe.
Many elaborate numerical codes have been constructed to model
the Universe on all scales, from dust agglomeration, over
planetary and stellar formation, to simulations of cosmological volumes.
Although these codes include a steadily increasing number of physical
processes, numerical resolution, etc., and although they are able to predict
various
observables, many of them fail to account for the fact that the observed light
may be rather different from the light that was emitted. Once
radiation leaves its origins, it may still be subject to various physical
processes altering not only its intensity, but in some cases also its
spatial and spectral distribution. If we
do not understand these processes, we may severely misinterpret the predictions
of the models when comparing to observations.
This is particularly true with regard to resonant lines like Ly$\alpha$, where
its complicated path may take it through regions of unknown physical conditions.
In order to interpret the observed Ly$\alpha$ lines correctly, it is crucial to
have an understanding of the physical processes that influence the shape and
transmission of the line.

Although the atomic processes regulating the individual scatterings are always
the same\footnote{As long the atom is not perturbed during the $\sim$10$^{-8}$
seconds
it is excited; this, however, can be safely ignored in most astrophysical
situations.}, the physical conditions governing different regions in space will
have a large impact on the outcome. For example, photons produced by
gravitational
cooling are mostly born in low-density regions, and may escape the galaxy more
or less freely. These photons thus tend to have a frequency close to the line
center. On the other hand, photons produced in the dense stellar regions will
usually have to scatter far from the line center in order to escape, and thus
tend to comprise the wings of the spectrum. However, these photons are also
more vulnerable to dust, since their paths are much longer,
and since also most of the galactic dust resides in these emission regions.
Moreover, the velocity field of the gas elements also has an effect on the
observed line profile.
In particular macroscopic gas motions, such as gas accretion or
outflows caused by feedback due to massive star formation,
will skew the line \citep[e.g.][]{dij06,ver06}.

While the history of theoretical Ly$\alpha$ radiative transfer (RT) dates back
to the 1930's
\citep{amb32,cha35}, the first practical prediction of spectra emerging
from a source embedded in neutral hydrogen was given by \citet{har73}, and later
generalized by \citet{neu90}. Due to the high opacity for a photon at the line
center, in general the photons have to diffuse in frequency and should hence
escape the medium in a double-peaked spectrum. Although this has indeed been
observed \citep[e.g.][]{yee91,ven05,van08}, apparently most Ly$\alpha$ profiles
from high-redshift galaxies seem to be missing the blue peak.
An immediate conclusion would be that high-redshift, Ly$\alpha$ emitting
galaxies (LAEs) are in the process of
massive star formation and thus exhibit strong outflows. Indeed, this scenario
has been invoked to explain a large number of LAE spectra, most convincingly
by \citet{ver08} who, assuming a central source and thin surrounding shell
of neutral gas while varying its expansion velocity, temperature, and gas and
dust column density, manage to produce nice fits to a number of observed
spectra.

Although the thin expanding shell scenario hinges on a physically plausible
mechanism, it is
obviously rather idealized. Furthermore, since most observations show only the
red peak of the profile, it seems to indicate that most (high-redshift)
galaxies exhibit outflows. However, at high redshifts many galaxies are still
forming, resulting in infall which would in
turn imply an increased \emph{blue} peak. Since this is rarely observed, in
this paper we aim to investigate whether some other mechanism, such as IGM
RT effects, could be responsible for
removing the blue peak and/or enhancing the red peak.

The present study complements other recent endeavors to achieve a
comprehensive understanding of how the Ly$\alpha$ line is redistributed in
frequency and real space \emph{after} having escaped its host galaxy
\citep{ili08,bar10,fau10,zhe10a,zhe10b,zhe10c}.
Our cosmological volume is not as large as
most of these studies, and although gas dynamics are included in the
simulations, ioniziation by UV radiation is calculated as a post-process rather
than on the fly.
The reward is a highly increased resolution, allowing
us to study the circumgalactic environs in great detail. Moreover, we
inquire into the temporal evolution of the IGM. 

The exploration of the high-redshift Ly$\alpha$ Universe is rapidly progressing,
gradually providing a census of the physical properties of young galaxies,
such as their luminosity functions, clustering properties, evolution, and
contents. Recent notable surveys have revealed hundreds of LAEs at high
redshifts, from $z \sim 2.1$--2.3 \citep{gua10,nil09},
to $z \sim 4.5$ \citep{wan09},
to $z \sim 6.6$ \citep{ouc10}, and for small samples even farther
\citep{hib10,til10}, and the present work, with its emphasis on the
reshaping of the Ly$\alpha$ line profile and amplitude throughout different
epochs is a step toward a more accurate interpretation of the observations.

In the following section, we discuss qualitatively the impact of the IGM on
the Ly$\alpha$ line profile. In Section \ref{sec:sims} we then decribe the
numerical simulations we have performed to study the effect quantitatively,
the results
of which are presented in Section \ref{sec:res}. We discuss our findings in
Section \ref{sec:disc}, and summarize in Section \ref{sec:sum}.


\section{The Intergalactic Medium}
\label{sec:igm}

As light travels through the
expanding Universe, it gets redshifted, implying that
wavelengths $\lambda$ blueward of the Ly$\alpha$ line center are eventually
shifted into resonance. If, for a given wavelength, this happens in the
vicinity of a sufficient amount of neutral
hydrogen, the spectrum experiences an absorption line (although strictly
speaking the
photons are not absorbed, but rather scattered out of the line of sight).
This results in the so-called Ly$\alpha$ forest \citep[LAF; ][]{lyn71,sar80}.

As the red part of the spectrum is only shifted farther away from resonance,
IGM absorption tends to skew
the line and not simply diminish it by some factor. In most earlier studies of
the line profiles of high-redshift galaxies, the IGM has either been
ignored \citep[e.g.][]{ver06,ver08},
taken to transmit the red half and remove the blue part \citep[e.g.][]{fin08},
or to influence the line uniformly by a factor
$e^{-\langle\tau\rangle}$, where $\langle\tau\rangle$ is the average optical
depth of the Universe \citep[e.g.][]{bru03,mei05,fau08,ryk09}.

Consider a source emitting the normalized spectrum
$\F_{\mathrm{em}}(\lambda) \equiv 1$ for all $\lambda$.
In an idealized, completely homogeneous universe undergoing completely
homologous expansion (``ideal Hubble flow'') and with the absorption profiles
having a negligible width, the observed spectrum would simply be a step
function, with $\Flam = \F_{\mathrm{red}} = 1$ for $\lambda > \lambda_0$, and
$\F(\lambda) = \F_{\mathrm{blue}} < 1$ for $\lambda \le \lambda_0$, where
$\lambda_0 = 1216$ {\AA} is the central wavelength of the line.
Three factors contribute to make $F$ differ from a step function:

Firstly, the Ly$\alpha$ line is not a delta function, but has a finite width,
in some cases extending significantly into the red part of the spectrum.
%
Secondly, the IGM is highly inhomogeneous, leading to large variations
in $\F$ blueward of $\lambda_0$ --- the LAF.
If we consider the average effect of the IGM and let $\F$ be the
average of many sightlines, $\F_{\mathrm{blue}}$ should still be a constant
function of wavelength. However,
in the proximity of galaxies the gas density is higher than far from
the galaxies; on the other hand, in these regions the stellar ionizing UV
radiation may reduce the neutral gas fraction.
Consequently, wavelengths just blueward of the line center may not on average
be subject to the same absorption as farther away from the line.
Regardless of which of
the two effects --- gas overdensity or stronger UV field --- is more important,
the correlation of the IGM with the source cannot be neglected.

Finally the expansion is not exactly homologous, since peculiar velocities of
the gas elements will cause fluctuations around the pure Hubble flow.
Considering again not individual sightlines but the average effect of the IGM,
these fluctuations are random and cancel out; on average, for every gas element
that recedes from a galaxy faster than the Hubble flow and thus causes an
absorption line at a slightly bluer wavelength, another gas element does the
opposite.
Hence, the average transmission in a ``realistic'' universe is the same as in a
universe where there are no peculiar velocities.
However, in the proximity of overdensities, the extra mass results in a
``retarded'' expansion of the local IGM.
When expansion around a source is somewhat slower than that of the rest
of the Universe, on average matter in a larger region will be capable of
causing absorption,
since the slower the expansion, the farther the photons will have to travel
before shifting out of resonance.

\subsection{The transmission function}
\label{sec:Fdef}

This paper focuses on two different aspects of the intergalactic
transmission. First, to see how the Ly$\alpha$ line is affected by the IGM, we 
calculate a ``transmission function'' given by the average, normalized flux
$\Flam$.
Specifically, $\F$ is calculated by taking the median value in
each wavelength bin of many sightlines, originating just outside a large number
of galaxies (where ``just outside'' will be defined later).
The standard deviation is defined by the 16 and 84 percentiles.
From the above discussion we may expect that $\F$ be characterized by a red
part $\F_{\mathrm{red}} \simeq 1$, and a blue part $\F_{\mathrm{blue}} < 1$,
but with a non-trivial shape just blueward of the line center, since the IGM in
the vicinity of the sources is different than far from the sources.


\subsection{The average transmission}
\label{sec:Tdef}

We will also examine the average transmission $\T = \T(z)$ of the
IGM, defined by first calculating for each individual sightline the fraction
of photons that are transmitted through the IGM in a relatively large
wavelength interval well away from the line center, and then taking the median
of all sightlines (again with the 16 and 84 percentiles defining the
standard deviation).
This quantity is sensitive to the overall ionization state of the IGM, and has
therefore been
used observationally to put constraints on the so-called \emph{Epoch of
Reionization} \citep[EoR; e.g.][]{bec01,djo01}.
A paramount puzzle in modern cosmology is the question of when and how the
hydrogen, and later helium, was reionized.
The EoR marks a comprehensive change
of the physical state of the gaseous Universe, and to understand the cause,
as well as the course, of this phenomenon is a challenging task.
Besides being a compelling event in itself, it also has profound
implications for the interpretation of observations and theoretical
cosmological models, not only due to the increased transparency of the
IGM, but also because of the accompanying rise in IGM temperature.

One way of probing the EoR is by looking at the spectra of high-redshift
quasars. The intrinsic spectrum of a quasar is characterized by
broad emission lines; in particular the Ly$\alpha$ line is very prominent.
For larger redshifts the absorption lines of the LAF gets
increasingly copious, until eventually
all of the light blueward of Ly$\alpha$ is absorbed, resulting in the so-called
Gunn-Peterson trough \citep{gun65}.
Observations of a large number of quasars
show that the Universe was largely opaque to radiation blueward of Ly$\alpha$
at $z \gtrsim 6$ \citep{son04,fan06}.

\section{Simulations}
\label{sec:sims}

\citet{lau09a,lau09b} considered numerically the shape of the Ly$\alpha$
spectrum emerging from galaxies at $z \sim 3.5$, based on hydro/gravity
galaxy simulations of very high resolution.
In these studies, the RT was carried out with Monte Carlo simulations,
following individual photons (or photon packets) as they scatter stochastically
out through the interstellar medium.
Since high resolution is important for the Ly$\alpha$ RT, especially when
taking into account dust, the simulations were conducted in galaxies extracted
from a fully cosmological, large volume model,
resimulated\footnote{A ``resimulation'' is performed by tracing the position of
the particles comprising a given galaxy back to the initial conditions, and
then repeating the simulation at 8 or 64 times the original mass resolution.}
at high resolution, and interpolated onto an adaptively refined grid.
This Ly$\alpha$ RT evidently only predicts the spectrum of
radiation that one would observe if located in the vicinity of the galaxies.
In reality the radiation has to travel trough the IGM. At a redshift
of $\sim$3.5, the IGM is largely ionized, and the spectra escaping the
galaxies are, in general, not very different from what would be observed at
Earth. However, even a very small amount of neutral hydrogen may influence the
observations.
To investigate just \emph{how} large an impact the IGM exerts on the radiation,
full IGM RT has to be computed. Furthermore, at higher redshifts where
the IGM is generally more neutral and more dense, neglecting the effect would
lead to severely erroneous results.

In principle this could be achieved by performing first the ``galactic'' RT in
the high-resolution resimulations and subsequently continuing the RT in the
low-resolution cosmological volume from the location of the individual
galaxies. However, although the physics of scattering in galaxies and that of
scattering in the IGM is not inherently different, the difference in physical
conditions imposes a natural division of the two schemes: in the dense gas of
galaxies, photons are continuously scattered in and out of the line of sight,
whereas in the IGM, once a photon is scattered out of the line of sight, it
is ``lost'', becoming part of the background radiation. The probability of a
background photon being scattered \emph{into} the line of sight, on the other
hand, is vanishingly small.

In order to disentangle galactic from intergalactic effects,
and, more importantly, to investigate the general effect of the IGM instead of
merely the IGM lying between us and a handful of resimulated galaxies, we take a
different approach: the transmission properties of the IGM are studied by
calculating the normalized spectrum $\Flam$ --- the \emph{transmission function}
--- in the vicinity of the Ly$\alpha$ line, as an average
of a large number of sightlines cast through the cosmological volume, and
originating just outside a large number of galaxies.

\subsection{Cosmological simulations}
\label{sec:cosm}

A standard, flat $\Lambda$CDM is assumed,
with $\Omega_m = 0.3$, $\Omega_\Lambda = 0.7$, and $h = H_0/100$ km $^{-1}$
Mpc$^{-1} = 0.7$.

The radiative transfer is carried out in a cosmological volume modeled
using self-consistent, \emph{ab initio} hydro/gravity simulations
\citep{som03,som06}.
The code used in these simulations is a significantly improved version of
the TreeSPH code, which has been used previously for galaxy formation
simulations \citep{som03}.
The main improvements over the previous version are:
(1) The ``conservative'' entropy equation solving scheme \citep{spr02} has
been implemented, improving shock resolution in SPH simulations.
In particular, this improves the resolution of
the starburst-driven galactic superwinds invoked in the simulations.
(2) Non-instantaneous gas recycling and chemical evolution, tracing the
ten elements H, He, C, N, O, Mg, Si, S, Ca and Fe \citep{lia02a,lia02b};
the algorithm invokes effects of
supernovae of type II and type Ia, and mass loss from stars of all masses.
(3) Atomic radiative cooling depending both on the metal abundance
of the gas and on the meta-galactic UV background (UVB), modeled after
\citet{haa96}, is invoked, as well as a simplified treatment
of radiative transfer, switching off the UVB where the gas
becomes optically thick to Lyman limit photons on a scale of 0.1 kpc
\citep{som06}.
The simple scheme for UVB RT is supplemented by a more elaborate
post-process ionizing UV RT scheme, as described in \Sec{uvrt}.

The simulation volumes were selected from two dark matter (DM) only
cosmological simulations of box length $10 h^{-1}$Mpc,
set up at $z_i=39$ using identical Fourier modes and phases, but with rms
linear fluctuations on a scale of $8 h^{-1}$Mpc, $\sigma_8$, equal to either
0.74 or 0.9, respectively, bracketing the latest
WMAP\footnote{{\tt http://lambda.gsfc.nasa.gov/product/map/current/}}-inferred
value of $\sim$0.8 \citep{jar10}.

The original DM-only simulations were run with periodic boundary
conditions using $128^3$ DM particles, and have been used previously
for galaxy resimulations \citep[e.g.][note, however, that in these simulations,
$\sigma_8=1.0$ was assumed]{som03,som06}.
For each of the two cosmological DM-only simulations,
mass and force resolution was increased (by factors of eight and two,
respectively) in a spherical Lagrangian sub-volume
of radius $5 h^{-1}$Mpc (about half the volume of the original DM-only
simulation), and in this region all DM particles were split into a DM particle
and a gas (SPH) particle according to an adopted universal baryon fraction of
$f_b=0.15$, in line with recent estimates.
The cosmological regions resimulated thus had a comoving volume
of about 1500 Mpc$^3$, contained about 17 million particles in total,
and were resimulated with an open boundary condition using the full
hydro/gravity code. The numerical resolution of these cosmological
resimulations was hence the same as that of the individual galaxy formation
simulations of \citet{som03}, reflecting the very considerable
increase in computing power since then.

The masses of SPH, star and DM particles were
$m_{\rm{gas}} = m_\star = 7.3\times10^5$ and
$m_{\rm{DM}}  = 4.2\times10^6$ $h^{-1}$ M$_{\odot}$, and the
gravity softening lengths were
$\epsilon_{\rm{gas}} = \epsilon_\star = 380$ and
$\epsilon_{\rm{DM}} = 680$ $h^{-1}$pc. The
gravity softening lengths were kept constant in physical coordinates from
$z=6$ to $z=0$, and constant in comoving coordinates at earlier times.

The choice of stellar initial mass function (IMF) affects the
chemical evolution and the energetics of stellar feedback processes,
in particular starburst-driven galactic superwinds.
Motivated by the findings of \citet{som08}, a \citet{sal55}
IMF was adopted for the simulations presented in this paper.


\subsection{Ionizing UV radiative transfer} 
\label{sec:uvrt}

The \citeauthor{haa96} UVB field becomes negligible at $z \ge \zre = 6$.
To comply with
the results of WMAP, which indicate a somewhat earlier onset of the ionizing
background, we ran an additional set of models, in which the UVB intensity
curve was ``stretched'' to initiate at $\zre = 10$.
In the following, Model 1, 2, and 3 refers to simulations using
$(\zre,\sigma_8) = (10,0.74)$, (10,0.9), and (6,0.74), respectively.
The models (also summarized in \Tab{models})
with $\zre = 10$ will be referred to as
 ``early'' reionization, while $\zre = 6$ will be referred to as ``late''
reionization.
In Sections \ref{sec:F} and \ref{sec:eor} we study the differences between
these models.

A realistic UV RT scheme \citep{raz06,raz07} was employed to post-process the
results of the hydrodynamical simulations using the following approach:
First, the physical properties of the SPH particles (density, temperature,
etc.) are interpolated onto an
adaptively refined grid of base resolution $128^3$, with dense cells
recursively subdivided into eight cells, until no cell contains more than ten
particles.
Around each stellar source, a number of radial rays are constructed and
followed through the nested grid, accumulating photoreaction number and 
energy rates in each cell.
Rays are split either as one moves farther away from the source, or as a
refined cell is entered.
These reaction rates are used to compute iteratively the equilibrium
ionization state of each cell.

In addition
to stellar photons, we also account for ionization and heating by LyC
photons originating outside the computational volume with the FTTE
scheme \citep{raz05} assuming the \citeauthor{haa96} UVB, modified to match
the particular reionization model.

Since the relative ratio of H{\sc i}, He{\sc i}, and He{\sc ii} densities
varies from
cell to cell, the UV RT is computed separately in three frequency bands,
$[13.6,24.6[$ eV, $[24.6,54.4[$ eV, and $[54.4,\infty[$ eV, respectively.
In each cell the angle-averaged intensity is added to the chemistry solver to
compute the ionization equilibrium.


\subsection{Galaxy selection criteria}
\label{sec:select}

Galaxies are located in the simulations as described in \citet{som05}.
To make sure that a given identified structure is a real galaxy, the
following selection criteria are imposed on the sample:
\begin{enumerate}
  \item To ensure that a given structure ``$i$'' is not just a substructure of
        a larger structure ``$j$'', if the center of structure $i$ is situated
        within the virial radius of $j$, it must have more stars than $j$.
  \item The minimum number of star particles must be at least
        $N_{\star\mathrm{,min}} = 15$. This corresponds to a minimum stellar
        mass of $\log(M_{\star\mathrm{,min}}/\Msun) = 7.2$.
  \item The circular velocity, given by
        $V_c = \sqrt{G M_{\mathrm{vir}} / r_{\mathrm{vir}}}$ must be
        $V_c \ge 35$ km s$^{-1}$.
\end{enumerate}

\begin{figure*}[!t]
\ifnum\astroph=0
  \epsscale{1.0}
\else
  \epsscale{1.2}
\fi
\plotone{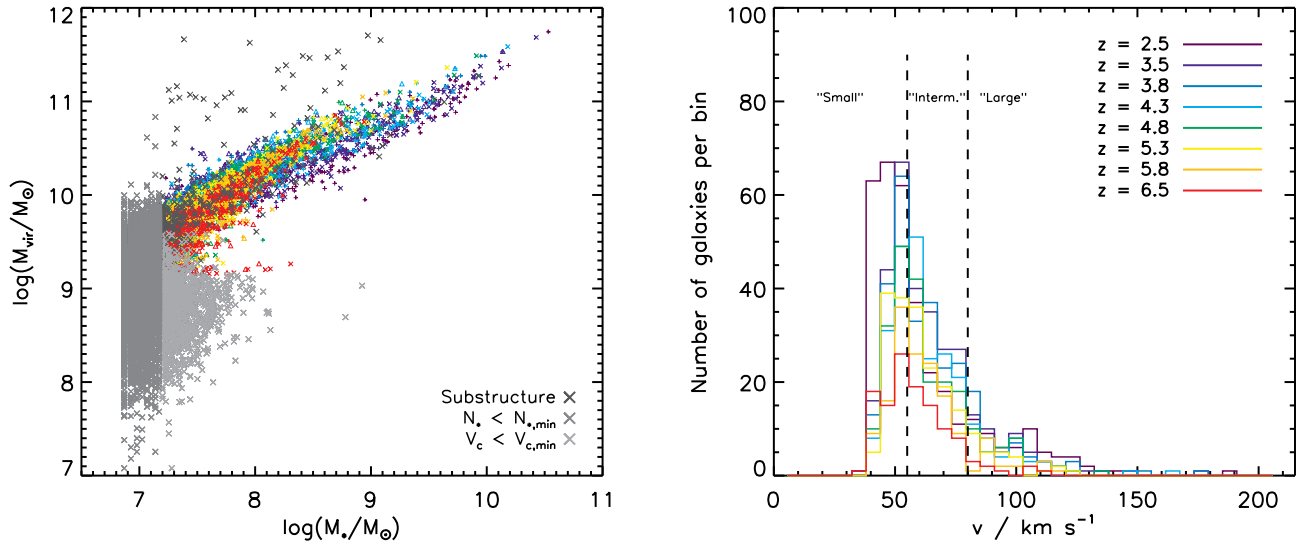}
\caption{\emph{Left:} Scatter plot of virial masses $M_{\mathrm{vir}}$
         vs.~stellar masses $M_\star$ for the full, unfiltered sample of
         galaxies. The colors signify redshift, with more red meaning higher
         redshift (exact values are seen in the right plot's legend).
         The data points of Model 1, 2, and 3 are shown with \emph{triangles},
         \emph{plus signs}, and \emph{crosses}, respectively.
         The galaxies that are rejected are overplottet with \emph{gray}
         crosses, with \emph{dark}, \emph{medium}, and \emph{light} gray
         corresponding to rejection criterion 1, 2, and 3, respectively.
         \emph{Right:} Distribution of circular velocities $V_c$ for the
         accepted galaxies in Model 1.
         The distributions for Model 2 and 3 look similar, although Model 2
         has more galaxies (see \Tab{models}).}
\label{fig:Filter}
\end{figure*}

The IGM in the vicinity of large galaxies is to some extend different from the
IGM around small galaxies. The more mass gives rise to a deeper gravitational
potential, enhancing the retarded Hubble flow. On the other hand, the larger
star formation may cause a larger bubble of ionized gas around it.
To investigate the difference in transmission, the sample of accepted galaxies
is divided into three subsamples, denoted ``small'', ``intermediate'', and
``large''. To use the same separating criteria at all redshifts, instead of
separating by mass --- which increases with time due to merging and accretion
--- the galaxies are separated according to their circular velocity, which
does not change significantly over time.
The thresholds are defined somewhat arbitrarily as $V_1 = 55$ km s$^{-1}$
(between small and intermediate galaxies) and $V_2 = 80$ km s$^{-1}$ (between
intermediate and large galaxies).
The final sample of galaxies is shown in \Fig{Filter}, which shows the relation
between stellar and virial mass, and the distribution of circular velocities.
\ifnum\astroph=0 \clearpage \fi

The exact number of galaxies in the three models studied in this work is seen
in \Tab{models}.
\begin{deluxetable}{cccccccc}
\tablecolumns{10}
\tablewidth{0pc}
\tablecaption{Number of galaxies in the simulations}
\tablehead{
\colhead{Model} &         $\zre$      &       $\sigma_8$      &  $z$  & Total & Small & Intermediate & Large \\
}                                                             
\startdata                                                    
          1.    &           10        &         0.74          &  2.5  &  343  &  207  &       80     &  56 \\
                &                     &                       &  3.5  &  309  &  138  &      127     &  44 \\
                &                     &                       &  3.8  &  283  &  125  &      119     &  39 \\
                &                     &                       &  4.3  &  252  &  102  &      115     &  35 \\
                &                     &                       &  4.8  &  225  &  111  &       86     &  28 \\
                &                     &                       &  5.3  &  201  &   98  &       85     &  18 \\
                &                     &                       &  5.8  &  154  &   75  &       62     &  17 \\
                &                     &                       &  6.5  &  121  &   70  &       44     &   7 \\
                                                                                                            \\
          2.    &           10        &         0.9           &  3.5  &  405  &  207  &      128     &  70 \\
                &                     &                       &  4.3  &  384  &  165  &      150     &  69 \\
                &                     &                       &  4.8  &  341  &  150  &      129     &  62 \\
                &                     &                       &  5.3  &  324  &  145  &      125     &  54 \\
                &                     &                       &  5.8  &  277  &  121  &      106     &  50 \\
                &                     &                       &  6.5  &  252  &  126  &       92     &  34 \\
                                                                                                            \\
          3.    &            6        &         0.74          &  3.5  &  325  &  162  &      122     &  41 \\
                &                     &                       &  3.8  &  318  &  165  &      117     &  36 \\
                &                     &                       &  4.3  &  293  &  150  &      110     &  33 \\
                &                     &                       &  4.8  &  250  &  138  &       84     &  28 \\
                &                     &                       &  5.3  &  204  &  101  &       85     &  18 \\
                &                     &                       &  5.8  &  160  &   75  &       67     &  18 \\
                &                     &                       &  6.5  &  126  &   69  &       50     &   7 \\
\enddata
\tablecomments{``Small'', ``intermediate'', and ``large'' galaxies are defined
               as having circular velocities
               $V_c < 55$ km s$^{-1}$,
               55 km s$^{-1} \le V_c < 80$ km s$^{-1}$, and
               $V_c \ge 80$ km s$^{-1}$, respectively.\\
               }
\label{tab:models}
\end{deluxetable}
\ifnum\astroph=0 \clearpage \fi


\subsection{IGM radiative transfer}
\label{sec:IGMRT}

The IGM RT is conducted using the code {\sc IGMtransfer}\footnote{The code can
be downloaded from the following URL:\\
\href{http://www.dark-cosmology.dk/~pela/IGMtransfer.html}
{\tt http://www.dark-cosmology.dk/\~{}pela/IGMtransfer.html}.}.
For the RT we use the same nested grid used for the ionizing
UV RT. The transmission properties of the IGM are studied by calculating the
normalized spectrum $\Flam$ in the vicinity of the Ly$\alpha$ line, as an
average of a high number of sightlines cast through the simulated cosmological
volume.

The resulting value of $\Flam$ at wavelength $\lambda$ for a given sightline is
\begin{equation}
\label{eq:F}
\Flam = e^{-\tau(\lambda)}.
\end{equation}
The optical depth $\tau$ is the sum of contributions from all the cells
encountered along the line of sight:
\begin{equation}
\label{eq:tau}
\tau(\lambda) = \sum_i^{\mathrm{cells}}
                n_{\textrm{{\scriptsize H}{\tiny \hspace{.1mm}I}},i}
                \,r_i
                \,\sigma(\lambda + \lambda v_{||,i}/c).
\end{equation}
Here,
$n_{\textrm{{\scriptsize H}{\tiny \hspace{.1mm}I}},i}$ is the density of
neutral hydrogen in the $i$'th cell,
$r_i$ is the distance covered in that particular cell,
$v_{||,i}$ is the velocity component of the cell along the line of sight,
and
$\sigma(\lambda)$ is the cross section of neutral hydrogen.
Due to the resonant nature of the transition, the largest contribution at a
given wavelength will arise from the cells the velocity of which corresponds
to shifting the wavelength close to resonance.

Although no formal definition of the transition from a galaxy to the IGM
exists, we have to settle on a definition of where to begin the sightlines,
i.e.~the distance $r_0$ from the center of a galaxy.
Observed Ly$\alpha$ profiles result from scattering processes in first the
galaxy and subseqently the IGM, and regardless of the chosen value of $r_0$,
for consistency ``galactic Ly$\alpha$ RT'' should be terminated at the same
value when coupling the two RT schemes.
In galactic Ly$\alpha$ RT, individual photons are traced, while in IGM RT we
only consider the photons that are not scattered out of the line of sight.
Hence the sightlines should begin where photons are mainly scattered \emph{out
of} the line of sight, and only a small fraction is scattered \emph{into}
the line of sight.
Since more neutral gas is associated with larger galaxies, as well as with
higher redshift, clearly $r_0$ depends on the galaxy and the epoch, but
measuring $r_0$ in units of the virial radius $r_{\mathrm{vir}}$
helps to compare the physical conditions around different galaxies.
In Section \ref{sec:F} we argue the a reasonable value of $r_0$ is
$1.5 r_{\mathrm{vir}}$, and show that the final results are only mildly
sensitive to the actual chosen value of $r_0$.

At a given redshift, $\Flam$ is calculated as the median in each wavelength
bin of $10^3$
sightlines from each galaxy in the sample (using $10^4$ sightlines produces
virtually identical results). The number of galaxies amounts to
several hundreds, and increases with time (cf.~\Tab{models}).
The simulated volume is a spherical region
of comoving diameter $D_{\mathrm{box}}$, but to avoid any spuriosities at the
edge of the region only a $0.9 D_{\mathrm{box}}$ sphere is used. Due to the
limited size of the volume, in order to perform the RT until a sufficiently
short wavelength is redshifted into resonance the sightlines are allowed to
``bounce'' within the sphere, such that a ray reaching the edge of the sphere
is re-``emitted'' back in a random angle in such a way that the total volume is
equally well sampled. Note however that in general the wavelength region close
to the line that is affected by the correlation of the IGM with the source is
reached well before the the first bouncing.

The normalized spectrum is emitted at rest wavelength in the reference frame of
the
center of mass of a galaxy, which in turn may have a peculiar velocity relative
to the cell at which it is centered. This spectrum is then Lorentz transformed
between the reference frames of the cells encountered along the line of sight.
Since the expansion of space is approximately homologous, each cell can be
perceived as lying
in the center of the simulation, and hence this bouncing scheme does not
introduce any bias, apart from reusing the same volume several times for a given
sightline. However, since the sightlines scatter around stochastically and thus
pierce a given region from various directions, no periodicities arise in the
calculated spectrum.

To probe the average transmission, the sightlines are propagated until the
wavelength 1080 {\AA} has been redshifted into resonance, corresponding to
$\Delta z \simeq 0.1$. In this case the sightlines bounce roughly 30 (20) times
at $z = 2.5$ (6.5).


\section{Results}
\label{sec:res}

In the following sections, Model 1 is taken to be the ``benchmark'' model,
while the others are discussed in \Sec{10vs6}.

\subsection{The Ly$\alpha$ transmission function}
\label{sec:F}

\begin{figure*}
\epsscale{1.0}
\plotone{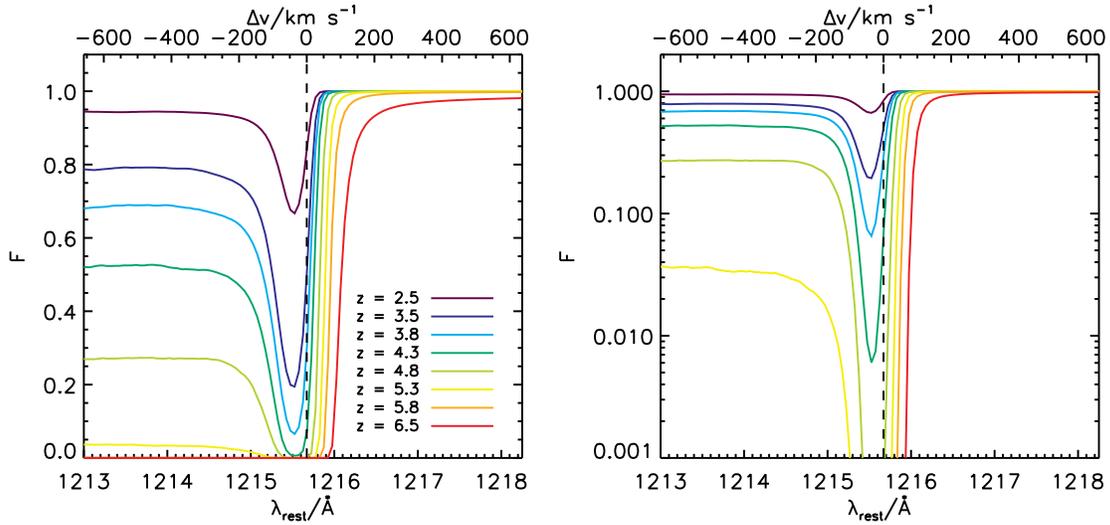}
\caption{Normalized transmission $\Flam$ at wavelengths around Ly$\alpha$,
         for different redshifts given by the color. In the \emph{left} panel,
         the vertical axis is linear, while in the \emph{right} it is
         logarithmic, emphasizing the transmission at high redshifts.}
\label{fig:Flam_z}
\end{figure*}

\begin{figure*}
\epsscale{1.0}
\plotone{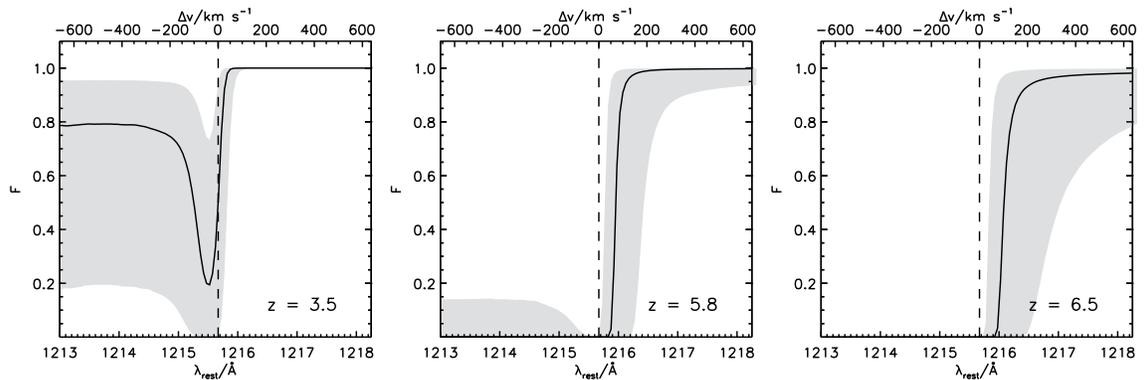}
\caption{Transmission $\F$ for $z = 3.5$ (\emph{left}), 5.8 (\emph{middle}),
         and 6.5 (\emph{right}) in Model 1, i.e. with $\sigma_8 = 0.74$ and
         $\zre = 10$. The shaded region indicates the
         range within which 68\% of the individual calculated transmission
         functions fall.}
\label{fig:Fscat}
\end{figure*}

\begin{figure*}
\epsscale{1.0}
\plotone{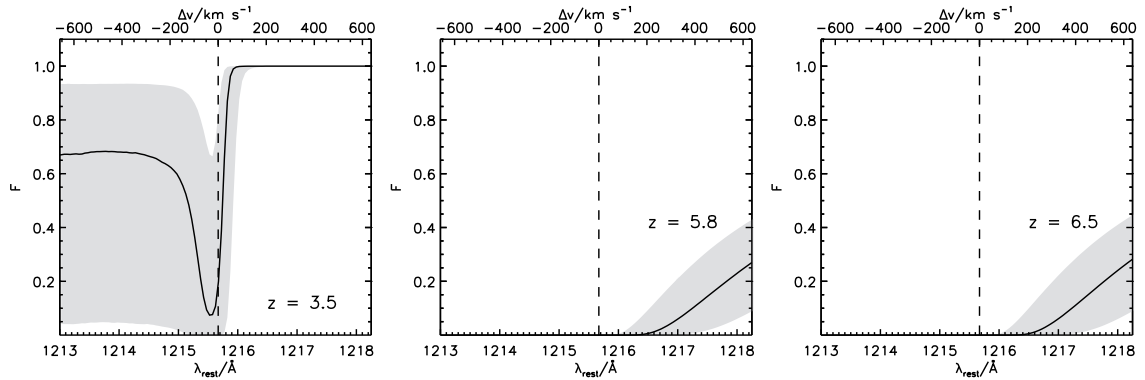}
\caption{Same as \Fig{Fscat} for Model 3, i.e. with $\zre = 6$. While at high
         redshifts a much more neutral IGM than in the $\zre = 10$ model causes
         a severe suppression of the Ly$\alpha$ line, by $z = 3.5$ the
         state of the IGM is not very different.}
\label{fig:Fscat_rz6}
\end{figure*}

Figure \ref{fig:Flam_z} shows the calculated transmission functions $\Flam$
at various redshifts.
Indeed, a dip just blueward of the line center is visible at
all redshifts.
The results in \Fig{Flam_z} were calculated as the median of sightines
emerging from all galaxies in the sample. Of course a large scatter exists,
since each sightline goes through very different regions, even if emanating
from the same galaxy.
\Fig{Fscat} shows the scatter associated with $\F$ for three different
redshifts. 
The equivalent transmission functions for the $\zre = 6$ model
are shown in \Fig{Fscat_rz6}. While at high redshifts the $\zre = 6$ model
clearly results in a much more opaque universe, at lower redshifts the
transmission properties of the IGM in the different models are more similar,
although at $z = 3.5$, the $\zre = 10$ model still transmits more light.

\begin{figure*}
\epsscale{1.0}
\plotone{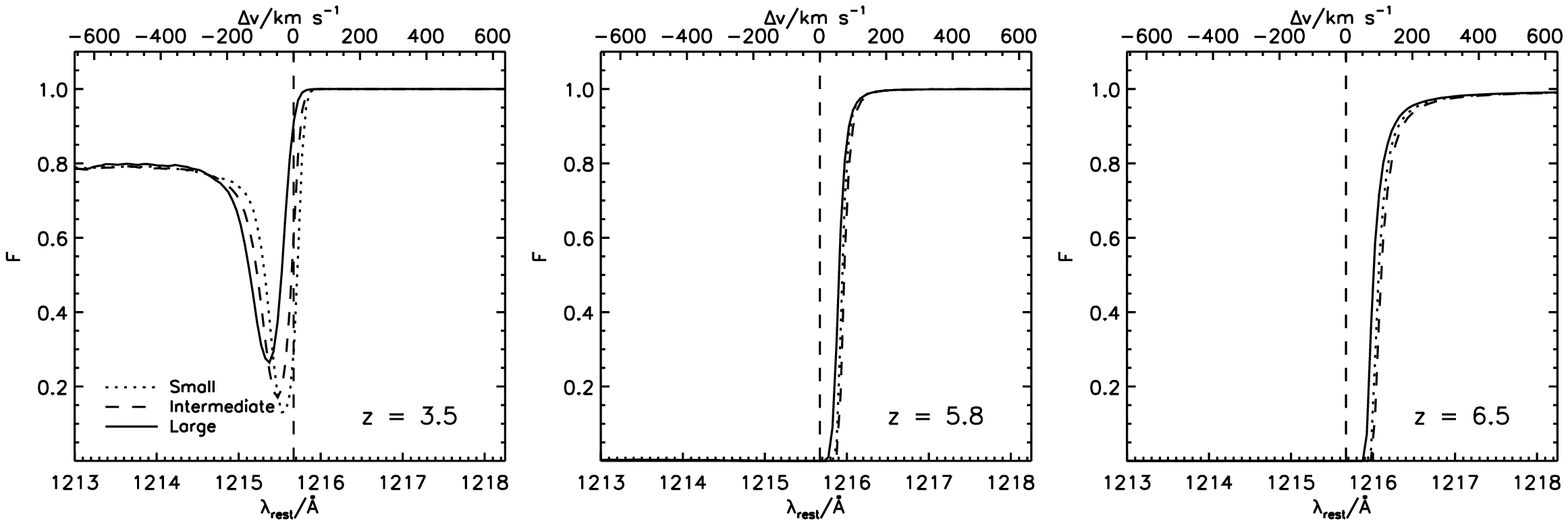}
\caption{Transmission $\F$ for $z = 3.5$ (\emph{left}), 5.8 (\emph{middle}), and
         6.5 (\emph{right}), for three different size categories of galaxies;
         small (\emph{dotted}),
         intermediate (\emph{dashed}), and
         large (\emph{solid lines}).
         Although slightly more absorption is seen in the vicinity of smaller
         galaxies, the transmission functions are quite similar for the three
         size ranges.}
\label{fig:LIS_z}
\end{figure*}

To see the difference in transmission around
galaxies of different sizes, \Fig{LIS_z}
shows transmission function for three size ranges (defined in \Sec{select}).
Since the distance from the galaxies at which the sightlines start is given in
terms of virial radii, sightlines emerging from small galaxies start closer to
their source than for large galaxies, and since at lower redshifts they tend
to be clustered together in the same overdensities as large galaxies, this
results in slightly more absorption. However, the difference is not very
significant.


\subsection{Effect on the spectrum and escape fraction}
\label{sec:eff}

\begin{figure*}
\ifnum\astroph=0
  \epsscale{1.0}
\else
  \epsscale{1.2}
\fi
\plotone{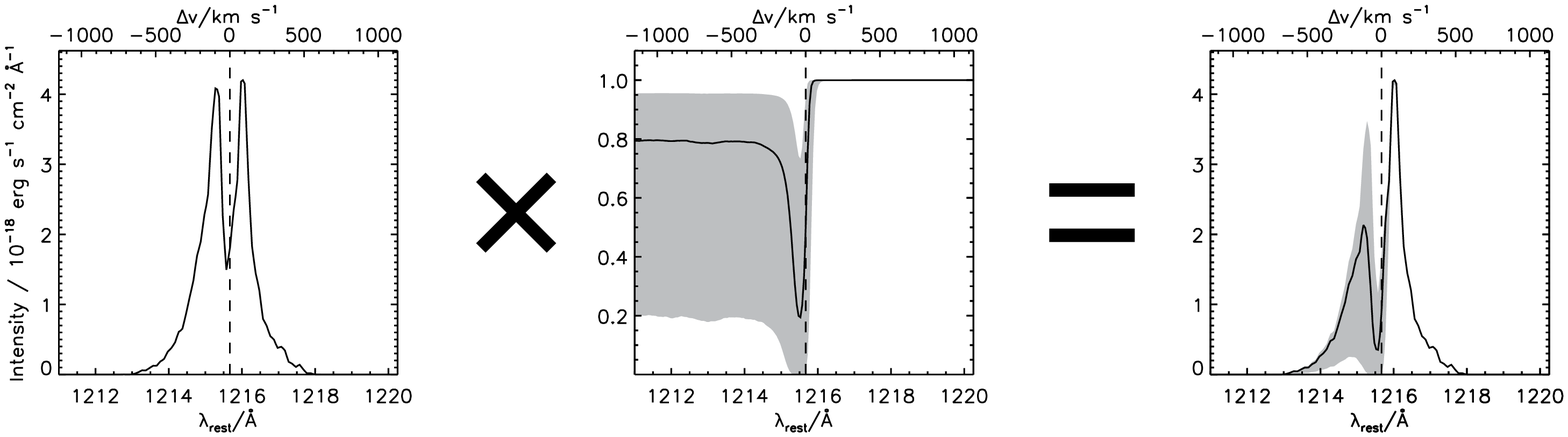}
\caption{Illustration of the effect of the IGM on the observed Ly$\alpha$
         profile emerging from a galaxy at $z \sim 3.5$. Without taking into
         account the IGM the two peaks are roughly equally high
         (\emph{left panel}). However, when the spectrum is transmitted through
         the IGM characterized by the transmission function $\Flam$
         (\emph{middle panel}), the blue peak is dimished, resulting in an
         observed spectrum with a higher red peak (\emph{right panel}).}
\label{fig:spXF}
\end{figure*}

In \Fig{spXF} the ``purpose'' of the transmission function is
illustrated: the left panel shows the spectrum emerging from a galaxy of
$M_{\mathrm{vir}} = 4.9 \times 10^9 \Msun$ and Ly$\alpha$ luminosity
$L_{\mathrm{Ly}\alpha} = 4.9 \times 10^{40}$ erg s$^{-1}$, calculated with the
Monte Carlo Ly$\alpha$ RT code {\sc MoCaLaTA} \citep{lau09a}.
Its circular velocity of 42 km s$^{-1}$ characterizes it as a small galaxy.
The spectrum which is
actually observed, after the light has been transferred through the IGM, is
shown in the right panel.
Although on average the effect of the IGM is not very large at this redshift,
as seen by the solid line in the right panel, due to the large dispersion in
transmission (visualized by the gray-shaded area) at least \emph{some} such
galaxies will be observed with a substantially diminished blue peak.
\ifnum\astroph=0 \clearpage \fi

\begin{figure*}
\epsscale{1.0}
\plotone{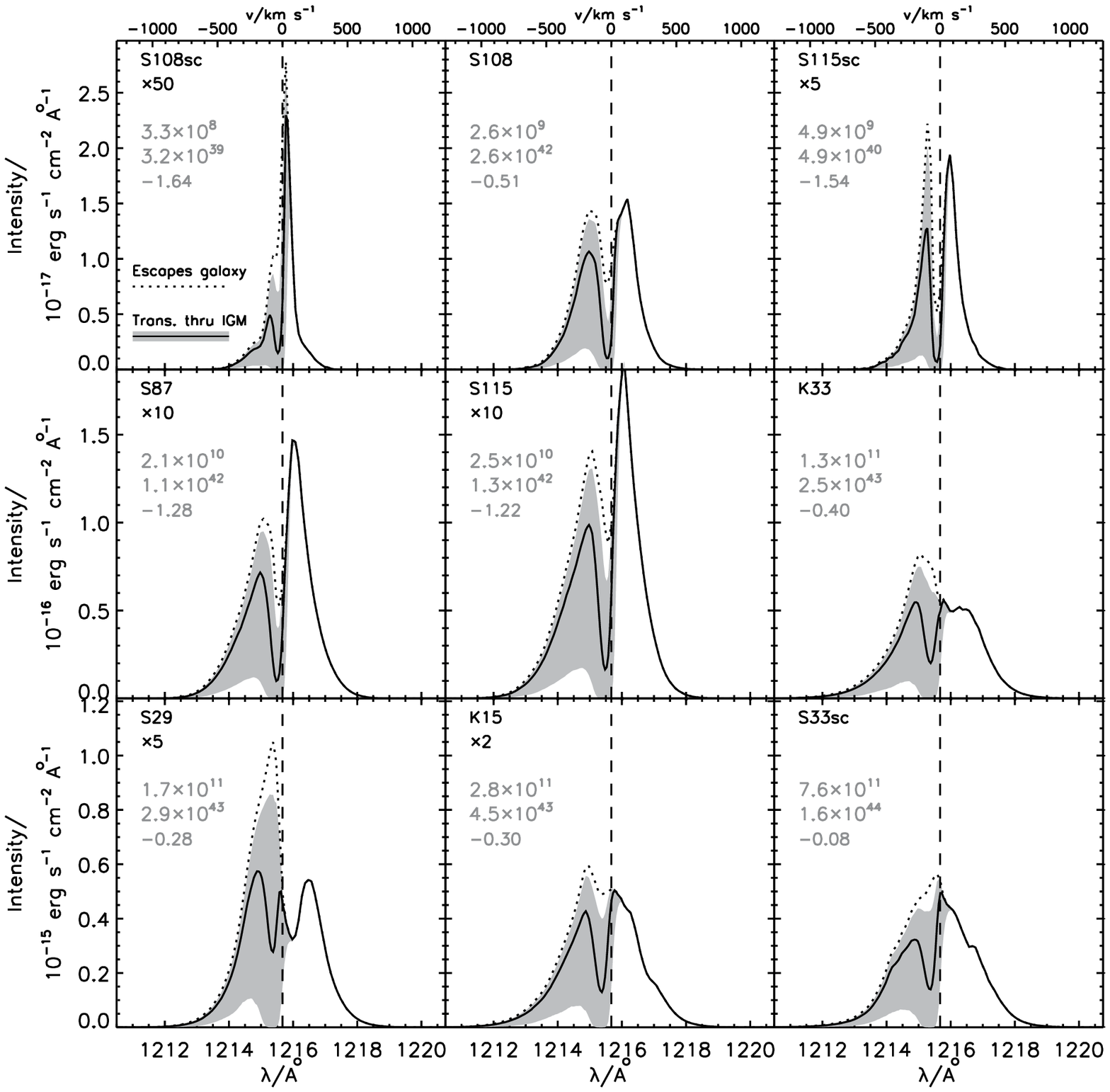}
\caption{Emitted spectra (\emph{dotted lines}) from nine different
         simulated galaxies at $\sim$3.5 --- ordered after increasing size ---
         and the corresponding spectra after being transmitted through the IGM
         (\emph{solid lines with gray regions denoting the 68\% confidence
         intervals}). The transmission functions appropriate for the given
         galaxy sizes have been used.
         In order to use the same ordinate axis for a given row, some
         intensities have been multiplied a factor indicated under the name of
         the galaxy.
         The numbers shown in gray are the corresponding galaxy's virial mass
         in $M_\odot$, its Ly$\alpha$ luminosity in erg s$^{-1}$, and its
         oxygen metallicity [O/H], respectively.}
\label{fig:spXFall}
\end{figure*}

In general, the larger a galaxy is the broader its emitted spectrum is, since
Ly$\alpha$ photons have to diffuse farther from the line center for higher
column densities of neutral gas. If dust is present, this will tend to narrow
the line \citep{lau09b}.
Larger galaxies tend to have higher metallicities and hence more dust, but
the lines will still be broader than the ones of small galaxies.
The galaxy used in \Fig{spXF} is quite small. In \Fig{spXFall} we show the
impact of the IGM on the nine simulated spectra from \citet{lau09b}, spanning
a range in stellar masses from $\sim$$10^7$ to $\sim$$10^{10}$.
For comparison, typical LAEs have stellar masses of
$M_\star/\Msun \sim 10^8$-$10^9$
\citep[][obtained from SED fitting]{gaw06,lai07,fin07,nil07}.

Besides altering the shape of the emitted spectrum, the IGM also has an effect
on another quantity of much interest to observers, namely the observed fraction
$f_{\mathrm{obs}}$ of the intrinsically emitted number of Ly$\alpha$ photons.
Since the bulk of the
emitted Ly$\alpha$ photons is due to young stars, the total Ly$\alpha$
luminosity of a galaxy may be used as a proxy for its star formation rate
(SFR), one of the main quantities characterizing galaxies. Assuming case B
recombination, $L_{\mathrm{Ly}\alpha}$ can be converted to a total H$\alpha$
luminosity $L_{\mathrm{H}\alpha}$
\citep[through $L_{\mathrm{H}\alpha} = L_{\mathrm{Ly}\alpha}/8.7$;][]{ost89},
which in turn can be converted to an SFR
\citep[through SFR = $7.9\times10^{-42} L_{\mathrm{H}\alpha}/$(erg s$^{-1}$)
 $\Msun$ yr$^{-1}$;][]{ken98}.

The above conversion factors assume that none of the emitted light is lost.
If dust is present in the galaxy a fraction of the emitted photons
will be absorbed
\citep[possibly making the LAE observable in FIR; see][]{day10a}.
This can be corrected for if the color excess $E(B-V)$ is
measured, assuming some standard extinction curve. However, this assumes that
both the H$\alpha$ and Ly$\alpha$ radiation are simply reduced by some factor
corresponding to having traveled the same distance through the dusty medium.
But since the path of Ly$\alpha$ photons is increased by resonant scattering,
this may be far from the truth. Comparing the Ly$\alpha$-inferred SFR with
that of H$\alpha$ (or UV continuum), the effect of scattering can be
constrained, as the quantity SFR(Ly$\alpha$)/SFR(H$\alpha$;UV) will be an
estimate of $f_{\mathrm{obs}}$.
Using this technique, values from a few percent
\citep[mostly in the nearby Universe; e.g.][]{hay07,ate08,hay10}
to $\sim$1/3 at high redshifts \citep{gro07} have been found.

Once the radiation enters the IGM, it is also not affected in the same way,
since the IGM is transparent to H$\alpha$, but not to Ly$\alpha$.
\citet{dij07} estimated analytically the fraction of Ly$\alpha$ photons
that are scattered out of the line of sight by the IGM (at $z \sim 4.5$--6.5)
and found a mean transmission of $f_{\mathrm{IGM}} \sim 0.1$--0.3, depending on
various assumptions (note that in \citet{dij07} this fraction is called
``$\mathcal{T}_\alpha$''). In that study, the intrinsic Ly$\alpha$ line profile
is modeled as a Gaussian, the width of which is given by the circular velocity
of the galaxies, in turn given by their mass, and assuming that no dust in the
galaxies alters the shape of the line before the light enters the IGM.

Since the transmission function is a non-trivial function of wavelength,
the exact shape and width of the Ly$\alpha$ line profile is important.
With the Ly$\alpha$ RT code {\sc MoCaLaTA} more realistic spectra can be
modeled, and applying the transmission function found in Section \ref{sec:F},
$f_{\mathrm{IGM}}$ can be calculated for the sample of simulated galaxies.
The transmitted fraction is not a strong function of galaxy size;
on average, a fraction $f_{\mathrm{IGM}} = 0.77_{-0.34}^{+0.17}$ of the photons
escaping the galaxies is transmitted through the IGM, at the investigated
redshift of $z \sim 3.5$.
The results for the individual galaxies is given in \Tab{fIGM}.
\begin{deluxetable}{lcccc}
\tablecolumns{5}
\tablewidth{0pc}
\tablecaption{Ly$\alpha$ transmission fractions for nine simulated galaxies at $z\sim3.5$}
\tablehead{
\colhead{Galaxy}  & $V_c$ & $f_{\mathrm{esc}}$ & $f_{\mathrm{IGM}}$              & $f_{\mathrm{obs}}$ \\
}                                                                                
\startdata                                                                       
         S108sc   &   17  &    $0.97\pm0.02$   & $0.66_{-0.36}^{+0.24}$          & $0.64_{-0.35}^{+0.23}$ \vspace{2mm}\\
         S108     &   33  &    $0.12\pm0.02$   & $0.78_{-0.35}^{+0.16}$          & $0.09_{-0.04}^{+0.02}$ \vspace{2mm}\\
         S115sc   &   42  &    $0.95\pm0.03$   & $0.76_{-0.34}^{+0.18}$          & $0.72_{-0.32}^{+0.17}$ \vspace{2mm}\\
         S87      &   69  &    $0.22\pm0.01$   & $0.81_{-0.32}^{+0.15}$          & $0.18_{-0.07}^{+0.03}$ \vspace{2mm}\\
         S115     &   73  &    $0.30\pm0.05$   & $0.80_{-0.34}^{+0.16}$          & $0.24_{-0.11}^{+0.06}$ \vspace{2mm}\\
         K33      &  126  &    $0.08\pm0.02$   & $0.78_{-0.34}^{+0.17}$          & $0.06_{-0.03}^{+0.02}$ \vspace{2mm}\\
         S29      &  137  &    $0.12\pm0.03$   & $0.75_{-0.35}^{+0.19}$          & $0.09_{-0.05}^{+0.03}$ \vspace{2mm}\\
         K15      &  164  &    $0.17\pm0.02$   & $0.79_{-0.37}^{+0.17}$          & $0.13_{-0.07}^{+0.03}$ \vspace{2mm}\\
         S33sc    &  228  &    $0.08\pm0.02$   & $0.80_{-0.33}^{+0.16}$          & $0.06_{-0.03}^{+0.02}$ \vspace{4mm}\\
         Average  &       &                    & $\mathbf{0.77_{-0.34}^{+0.17}}$ &                                    \\
\enddata
\tablecomments{Columns are, from left to right:
               galaxy name,
               circular velocity $V_c$ in km s$^{-1}$,
               fraction $f_{\mathrm{esc}}$ of emitted photons escaping the
                 galaxy (i.e.~not absorbed by dust),
               fraction $f_{\mathrm{IGM}}$ of these transmitted through the
                 IGM, and
               resulting observed fraction
                 $f_{\mathrm{obs}} = f_{\mathrm{esc}} f_{\mathrm{IGM}}$.
               Uncertainties in $f_{\mathrm{esc}}$ represent varying escape
               fractions in different directions, while uncertainties in
               $f_{\mathrm{IGM}}$ represent variance in the IGM.
               Uncertainties in $f_{\mathrm{obs}}$ are calculated as
               $(\sigma_{\mathrm{obs}} / f_{\mathrm{obs}})^2 =
                (\sigma_{\mathrm{esc}} / f_{\mathrm{esc}})^2 +
                (\sigma_{\mathrm{IGM}} / f_{\mathrm{IGM}})^2$.}
\label{tab:fIGM}
\end{deluxetable}

\begin{figure*}
\epsscale{1.0}
\plotone{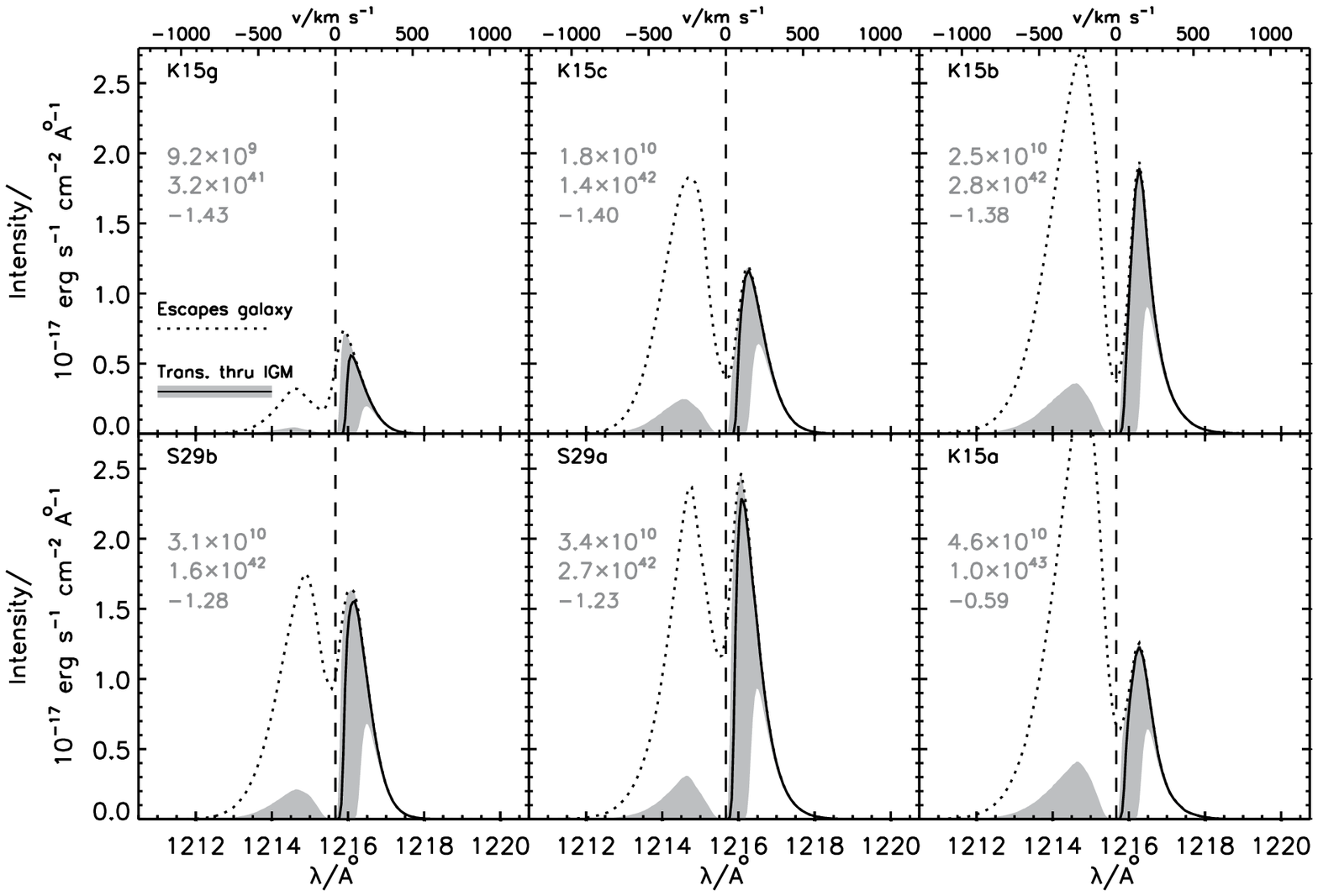}
\caption{Same as \Fig{spXFall}, but for $z = 5.8$.}
\label{fig:spXFall_z5p8}
\end{figure*}

\begin{figure*}
\epsscale{1.0}
\plotone{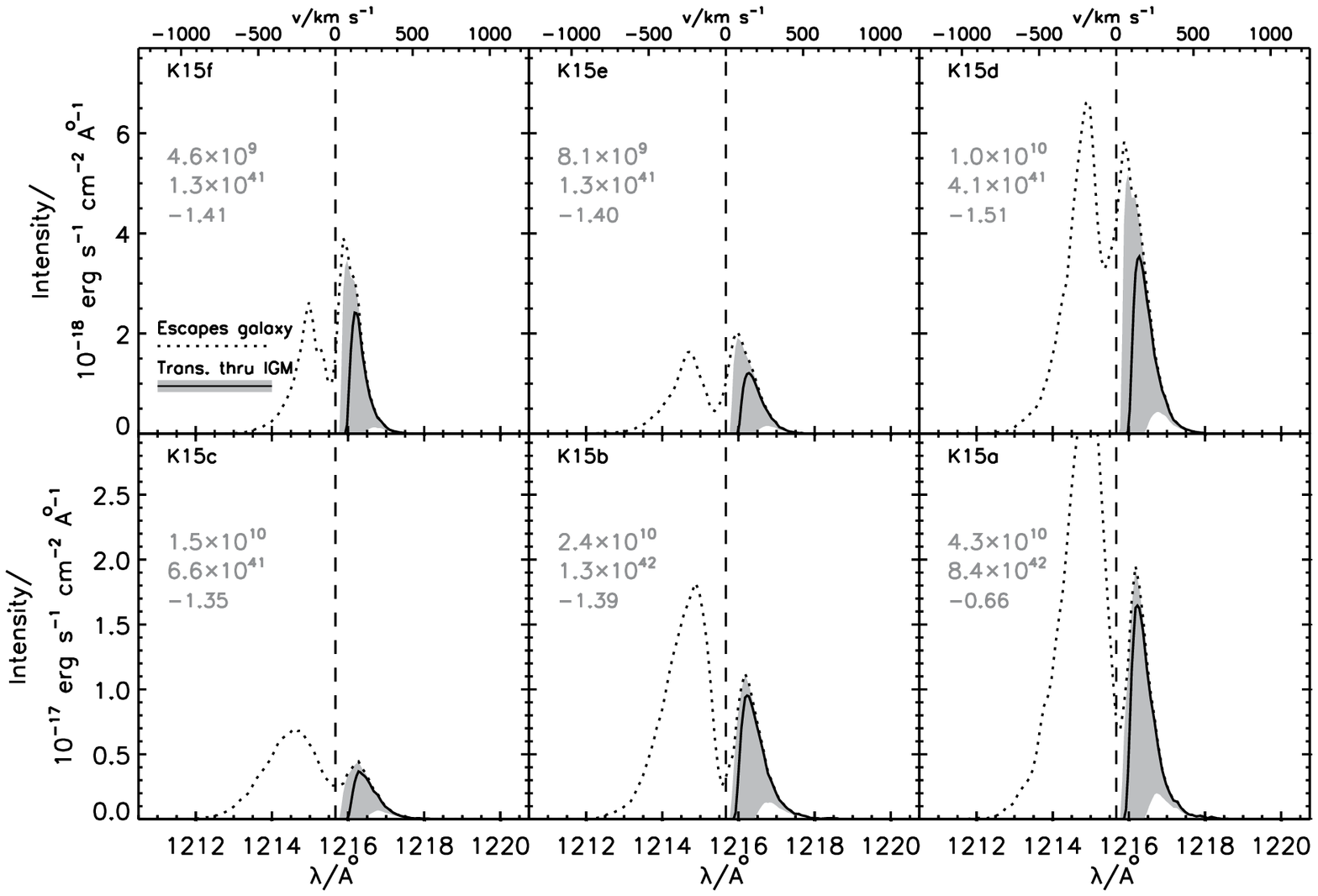}
\caption{Same as \Fig{spXFall}, but for $z = 6.5$.}
\label{fig:spXFall_z6p5}
\end{figure*}

At higher redshifts the metallicity is generally lower,
leading to less dust and
hence larger values of $f_{\mathrm{esc}}$. However, the increased neutral
fraction of the IGM scatters a correspondingly higher number of photons out of
the line of sight, resulting in a smaller total observed fraction.
\Fig{spXFall_z5p8} and \Fig{spXFall_z6p5} shows the impact of the IGM on
Ly$\alpha$ profiles at $z \sim 5.8$ and $z \sim 6.5$, respectively,
and \Tab{fIGM_z5p8} and \Tab{fIGM_z6p5} summarizes the obtained fractions.
At these redshifts, for the six galaxies for which the calculations have been
carried out on average a fraction
$f_{\mathrm{IGM}}(z=5.8) = 0.26_{-0.18}^{+0.13}$ and
$f_{\mathrm{IGM}}(z=6.5) = 0.20_{-0.18}^{+0.12}$ of the photons is transmitted
through the IGM, consistent with what was obtained by \citet{dij07}.

\begin{deluxetable}{lcccc}
\tablecolumns{5}
\tablewidth{0pc}
\tablecaption{Ly$\alpha$ transmission fractions for six simulated galaxies at $z\sim5.8$}
\tablehead{
\colhead{Galaxy}  & $V_c$ & $f_{\mathrm{esc}}$ & $f_{\mathrm{IGM}}$              & $f_{\mathrm{obs}}$ \\
}                                                                                
\startdata                                                                       
         K15g     &   63  &    $0.93\pm0.02$   & $0.35_{-0.24}^{+0.16}$          & $0.33_{-0.23}^{+0.15}$ \vspace{2mm}\\
         K15c     &   78  &    $0.82\pm0.02$   & $0.25_{-0.14}^{+0.11}$          & $0.21_{-0.11}^{+0.09}$ \vspace{2mm}\\
         K15b     &   88  &    $0.46\pm0.02$   & $0.23_{-0.14}^{+0.12}$          & $0.11_{-0.06}^{+0.05}$ \vspace{2mm}\\
         S29b     &   94  &    $0.85\pm0.01$   & $0.27_{-0.17}^{+0.14}$          & $0.23_{-0.15}^{+0.12}$ \vspace{2mm}\\
         S29a     &   96  &    $0.52\pm0.08$   & $0.31_{-0.22}^{+0.14}$          & $0.16_{-0.12}^{+0.07}$ \vspace{2mm}\\
         K15a     &  108  &    $0.17\pm0.05$   & $0.16_{-0.11}^{+0.11}$          & $0.03_{-0.02}^{+0.02}$ \vspace{4mm}\\
         Average  &       &                    & $\mathbf{0.26_{-0.18}^{+0.13}}$ &                                    \\
\enddata
\tablecomments{Same as \Tab{fIGM}, but for $z = 5.8$.}
\label{tab:fIGM_z5p8}
\end{deluxetable}
\begin{deluxetable}{lcccc}
\tablecolumns{5}
\tablewidth{0pc}
\tablecaption{Ly$\alpha$ transmission fractions for six simulated galaxies at $z\sim6.5$}
\tablehead{
\colhead{Galaxy}  & $V_c$ & $f_{\mathrm{esc}}$ & $f_{\mathrm{IGM}}$              & $f_{\mathrm{obs}}$ \\
}                                                                                
\startdata                                                                       
         K15f     &   50  &    $0.96\pm0.03$   & $0.24_{-0.22}^{+0.18}$          & $0.23_{-0.22}^{+0.17}$ \vspace{2mm}\\
         K15e     &   62  &    $0.91\pm0.04$   & $0.24_{-0.22}^{+0.14}$          & $0.22_{-0.20}^{+0.13}$ \vspace{2mm}\\
         K15d     &   66  &    $0.95\pm0.02$   & $0.18_{-0.16}^{+0.13}$          & $0.17_{-0.16}^{+0.12}$ \vspace{2mm}\\
         K15c     &   76  &    $0.88\pm0.02$   & $0.16_{-0.14}^{+0.07}$          & $0.14_{-0.13}^{+0.06}$ \vspace{2mm}\\
         K15b     &   89  &    $0.83\pm0.03$   & $0.20_{-0.17}^{+0.06}$          & $0.17_{-0.14}^{+0.05}$ \vspace{2mm}\\
         K15a     &  108  &    $0.26\pm0.09$   & $0.18_{-0.16}^{+0.06}$          & $0.05_{-0.04}^{+0.02}$ \vspace{4mm}\\
         Average  &       &                    & $\mathbf{0.20_{-0.18}^{+0.12}}$ &                                    \\
\enddata
\tablecomments{Same as \Tab{fIGM}, but for $z = 6.5$.}
\label{tab:fIGM_z6p5}
\end{deluxetable}
%


\subsection{Probing the Epoch of Reionization}
\label{sec:eor}

We now focus on a different topic, namely the RT in the IGM far from the
emitting galaxies.
Measuring the average transmission in a wavelength interval blueward of the
Ly$\alpha$ line for a sample of quasars or other bright sources, one can
ascertain the average
transmission properties and hence the physical state of the IGM.
The interval in which the transmission is calculated should be large
enough to achieve good statistics, but short enough that
the bluest wavelength does not correspond to a redshift epoch appreciably
different from the reddest.
Furthermore, in order to probe the real IGM and not the quasar's neighborhood,
the upper limit of the wavelength range should be taken to have a value
somewhat below $\lambda_0$.

\begin{figure}
\ifnum\astroph=0
  \epsscale{1.0}
\else
  \epsscale{1.2}
\fi
\plotone{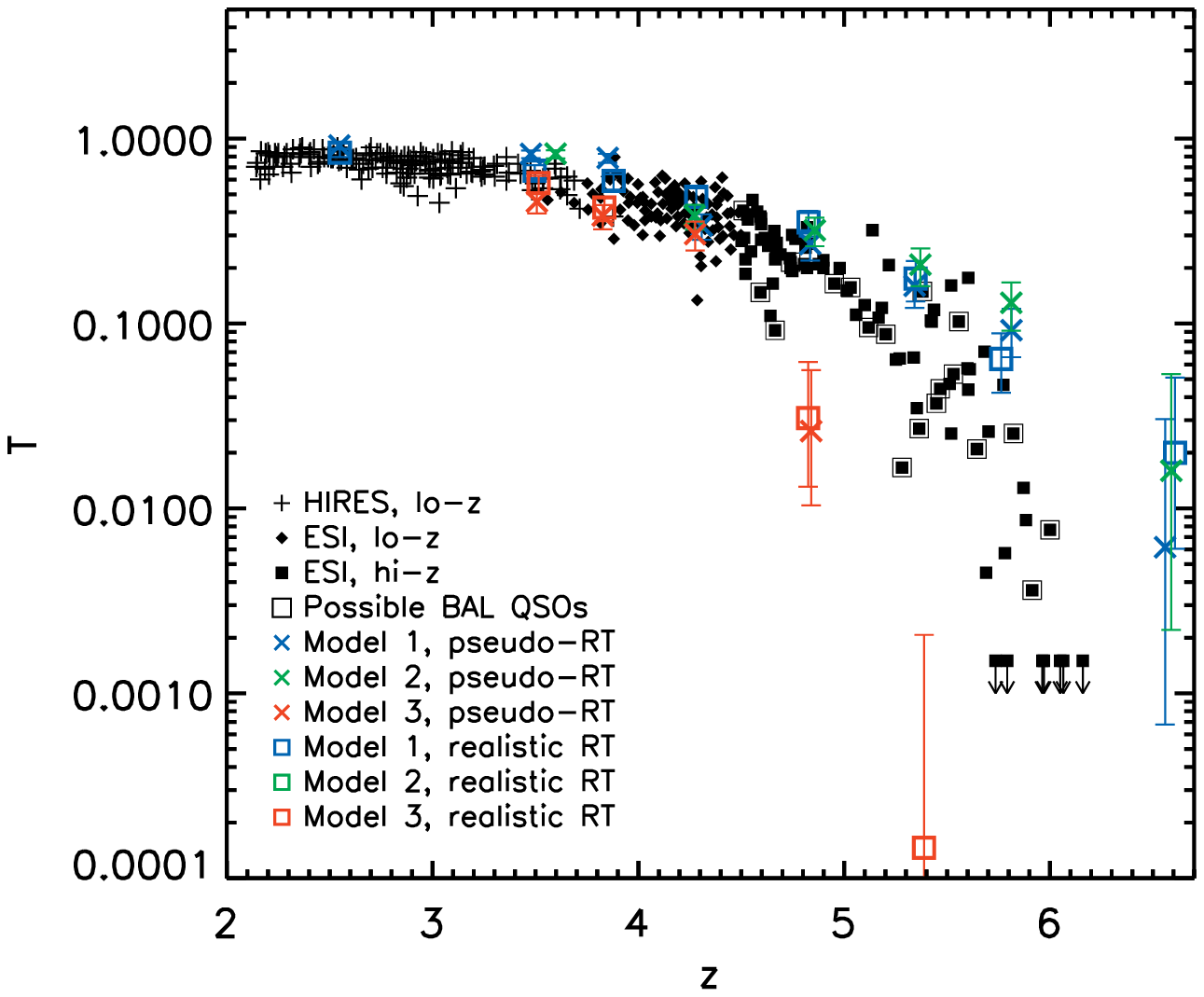}
\caption{Comparison of observations (\emph{black data points}) and simulations
         (\emph{colored data points}) of the transmitted flux blueward of the
         Ly$\alpha$ line as a function of redshift.
         The three models are represented by the colors \emph{blue},
         \emph{green}, and \emph{red} for model 1, 2, and 3, respectively.
         To highlight the significance of the improved UV RT, we show both the
         transmission in the ``original'' simulation with the ``pseudo''-RT
         (\emph{crosses}) and with the improved UV RT (\emph{squares}).
         For details on the observations see \citet{son04}, from where the data
         are kindly supplied.}
\label{fig:Songaila}
\end{figure}

\citet{son04} measured the IGM transmission in the LAFs of a large sample of
quasars with redshifts between 2 and 6.5, in the wavelength interval
1080--1185 {\AA}. In \Fig{Songaila} we
compare the simulated transmitted fractions with her sample.
As is evident from the figure, an ``early'' reionization, i.e. with $\zre = 10$
(Model 1 and 2),
yields a slightly too transparent Universe, while a ``late'' reionization
(Model 3)
yields a too opaque Universe. Note that the log scale makes the
$\zre = 6$ data seem much farther off than the $\zre = 10$ data. 


\subsection{Convergence test}
\label{sec:convtest}

\begin{figure*}
\epsscale{1.0}
\plotone{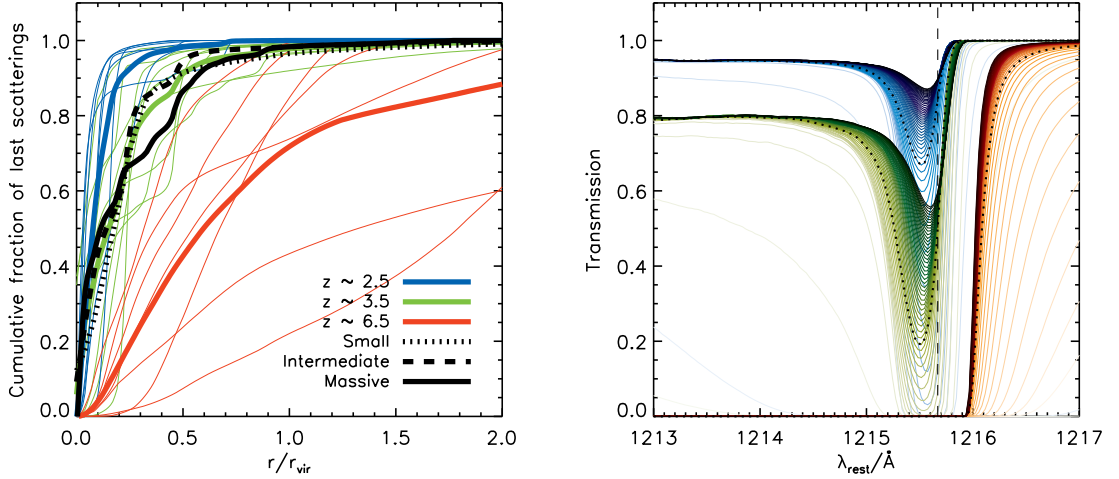}
\caption{\emph{Left:}
         Cumulative probability distribution of the distance $r$ from the
         center of a galaxy at which the last scattering takes place
         (calculated with the Ly$\alpha$ RT code {\sc MoCaLaTA} \citep{lau09a}).
         \emph{Thin solid} lines represent individual galaxies at redshift
         2.5 (\emph{blue}), 3.5 (\emph{green}), and 6.5 (\emph{red}),
         while \emph{thick solid colored} lines are the average of these.
         Also shown, in \emph{black}, are the average of three different size
         ranges at $z = 3.5$;
         small (\emph{dotted}), intermediate (\emph{dashed}), and large
         galaxies (\emph{solid}).
         \ifnum\astroph=1 \\ \fi
         \emph{Right:} Resulting transmission function $\Flam$ for sightlines
         originating at various distances $r_0/r_{\mathrm{vir}}$ from galactic
         centers, increasing in steps of $0.1 r_{\mathrm{vir}}$ and ranging
         from $r_0 = 0.1 r_{\mathrm{vir}}$ to $r_0 = 5 r_{\mathrm{vir}}$.
         The three different redshifts are shown in shades of \emph{blue},
         \emph{green}, and \emph{red}, for $z = 2.5$, 3.5, and 6.5,
         respectively, and the lines go from light shades for low values of
         $r_0$ to dark shades for high values of $r_0$.
         The transmission functions corresponding to
         $r_0 = 1.5 r_{\mathrm{vir}}$ are shown in \emph{black dotted} lines.
         All results are for Model 1.}
\label{fig:X_i}
\end{figure*}

In Section \ref{sec:IGMRT} we stated that the sightlines should initiate at
a distance $r_0$ from the centers of the galaxies given by their virial radius.
Figure \ref{fig:X_i} displays the cumulative probability
distributions of the distance of the last scattering from the galaxy center
(calculated with {\sc MoCaLaTA}) for
different redshifts and mass ranges, demonstrating that in most cases a photon
will have experienced its last scattering at the order of 1 $r_{\mathrm{vir}}$
from the center of its host
galaxy. For increasing redshift, the photons tend to escape the galaxies at
larger distances due to the higher fraction of neutral hydrogen, but the change
with redshift seems quite slow. Furthermore, at a given redshift the
distinction between different galactic size ranges appears insignificant
(as long as $r_0$ is measured in terms of $r_{\mathrm{vir}}$).

Also shown in \Fig{X_i} are the transmission curves for sightlines
initiating at various distances $r_0$ from the centers of the galaxies.
For very small values of $r_0$, a significantly lower
fraction is transmitted due to the high density of neutral gas. Around
$r \sim r_{\mathrm{vir}}$
the change in $\Flam$ becomes slow, converging to $\Flam$ exhibiting no dip
for $r_0 \to \infty$.

In summary, scaling $r_0$ to the virial radius $r_{\mathrm{vir}}$ of the
galaxies allows us to use the same value of $r_0 = 1.5 r_{\mathrm{vir}}$ for
all sightlines.



\section{Discussion}
\label{sec:disc}

\subsection{The origin of the dip}
\label{sec:dip}

In general, the effect of the IGM --- \emph{even at relatively low redshifts}
--- is to reduce the blue peak of the Ly$\alpha$ line profile.
At $z\sim3.5$, the effect is not strong enough to fully
explain the oft-observed asymmetry,
but at $z \gtrsim 5$, almost all radiation blueward of the line center
is lost in the IGM.

\begin{figure*}
\epsscale{1.1}
\plotone{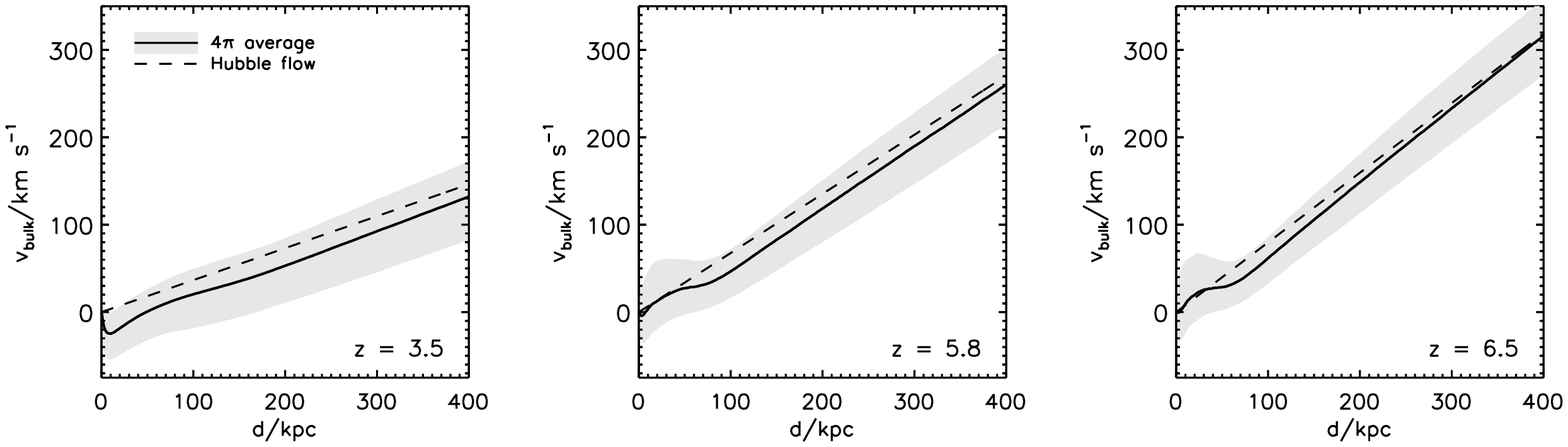}
\caption{Average recession velocity $v_{\mathrm{bulk}}$ of the IGM as a
         function of proper distance $d$ from the centers of the galaxies in
         Model 1 (\emph{solid black} lines, with \emph{gray} regions indicating
         the 68\% confidence intervals).
         At all redshift, the expansion is retarded compared to the pure Hubble
         flow (\emph{dashed} line) out to a distance of several comoving Mpc.
         At high redshifts,
         however, very close to the galaxies outflows generate higher recession
         velocities.}
\label{fig:probeV}
\end{figure*}

\begin{figure*}
\epsscale{1.1}
\plotone{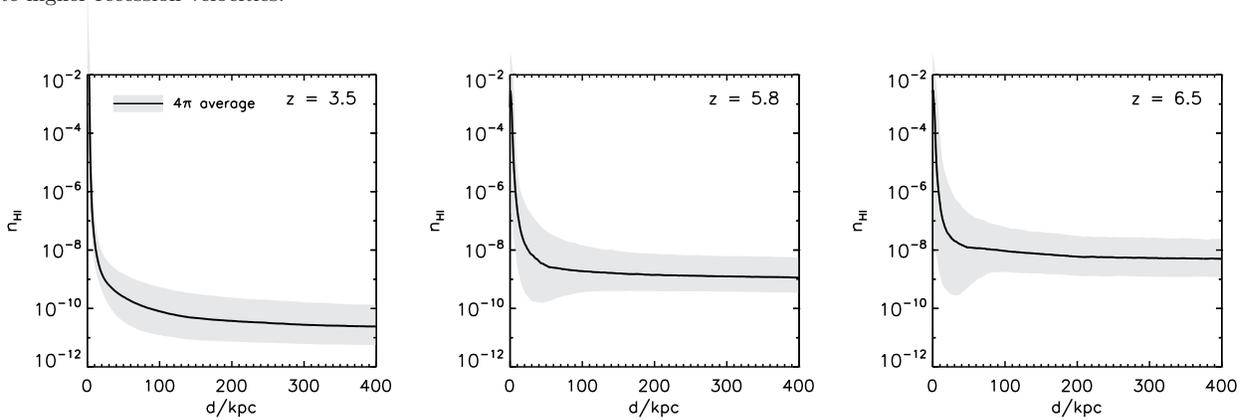}
\caption{Average density $\nhi$ of neutral hydrogen as a
         function of proper distance $d$ from the centers of the galaxies in
         Model 1 (\emph{solid black} lines, with \emph{gray} regions indicating 
         the 68\% confidence intervals).
         While in general the density decreases with distance, at high
         redshifts ionizing radiation reduces $\nhi$ in the immediate
         surroundings of at least some of the galaxies, as seen by the small
         dip at $\sim$50 kpc in the lower part of the gray zone.}
\label{fig:probenHI}
\end{figure*}

Figures \ref{fig:probeV} and \ref{fig:probenHI} display for three different
epochs the velocity field and the density of neutral hydrogen, respectively,
of the IGM associated with the galaxies, taken as an average of all galaxies in
the samples, and in all directions.
At all redshifts, on average the IGM close to the galaxies recedes at a
somewhat slower rate than that given by the pure Hubble flow, slowly converging
toward the average expansion rate of the Universe.
At the higher
redshifts, the IGM in the immediate proximity of the galaxies is characterized
by higher velocities, due to outflows generated by starbursts. At $z = 3.5$,
however, this effect is overcome by the accretion of gas.

Inspecting the dips in \Fig{Flam_z}, the minima are seen to be located at
roughly 50 km s$^{-1}$, almost independently of the redshift but becoming
slightly broader with increasing $z$.
At $z = 3.5$, the dip extends all the way out to 300--400 km s$^{-1}$.
This corresponds to the central absorption being caused by the IGM within
$\sim$150 kpc, and the wings of the absorption by the IGM within $\sim$1 Mpc.
As shown in \Fig{probenHI}, at $z = 3.5$ the density of neutral hydrogen
decreases monotonically with distance from the source, and within $\sim$150
kpc $\nhi$ is substantially higher than the cosmic mean. Further away, the
density is close to the mean density of the Universe. However, as seen in
\Fig{probeV} the recession velocity of the gas continues to lie below that of
the average, Universal expansion rate, and in fact does so until approximately
1 Mpc from the source.\footnote{Since the cosmological volume is several
Mpc across, except for the galaxies
lying close to the edge, the absorption takes place before the sightlines
``bounce'', so in general there is no risk of a sightline going through the
same region of space before having escaped the zone causing the dip.}

Thus, the cause of the suppression of the blue wing of the Ly$\alpha$ line may,
at wavelengths close to the line center ($\Delta\lambda \simeq 1/2$ {\AA}),
be attributed to an increased density of neutral hydrogen close to the galaxies,
while farther away from the line center ($\Delta\lambda \lesssim$ 1--1.5 {\AA}),
to a retarded Hubble flow.

\citet{zhe10a} perform Ly$\alpha$ RT in the IGM at $z = 5.7$ through full Monte
Carlo
simulations, and make a detailed comparison with the observed line profile
obtained from simply multiplying the intrinsic line by $e^{-\tau}$, as has
been done in previous models \citep[e.g.][]{ili08}. They conclude that
neglecting scattering effects severely underestimates the transmitted fraction.
While it is certainly true that treating scattering processes as absorption
inside the galaxies is only a crude approximation, once the probability of
photons being scattered \emph{into} the line of sight becomes sufficiently
small, this approach is quite valid. In their analysis, \citet{zhe10a} start
their photons in the center of the galaxies, which are resolved only by a few
cells (their $dx$ being $\sim$28 kpc in physical coordinates and their
fiducial galaxy having $r_{\mathrm{vir}}$ = 26 kpc).
Since the side length of our smallest cells are more than 400
times smaller than the
resolution of \citet{zhe10a}, we are able to resolve the galaxies and their
surroundings in great detail, and we are hence able to determine the distance
at which the $e^{-\tau}$ model becomes realistic.
Moreover, when coupling the IGM RT with Ly$\alpha$ profiles, we use the
realistically calculated profile, whereas \citet{zhe10a} use a Gaussian set
by the galaxies' halo masses. Their line widths $\sigma_{\mathrm{init}} =
32 M_{\mathrm{10}}^{1/3}$ km s$^{-1}$,
 where $M_{\mathrm{10}}$ is the halo mass divided by
$10^{10} h^{-1} M_\odot$, thus neglect broadening by scattering.
This makes them much smaller than ours, which are typically
several hundred km s$^{-1}$.
Note, however, that our relatively small cosmological volume and the fact that
ionizing UV RT is performed as a post-process rather than on the fly may make
our density field less accurate than that of \citet{zhe10a}.

The photons that are scattered out of the line of sight are of course not lost,
but rather become part of a diffuse Ly$\alpha$ background. Since more
scatterings take place in the vicinity of galaxies, LAEs tend to be
surrounded by a low-surface brightness halo, making them look more extended on
the sky when comparing to continuum bands on the $r \sim 10$--100 kpc scale.
This has been confirmed observationally
\citep[e.g.][but see \citet{bon10}]{mol98,fyn01,fyn03}, and studied
theoretically/numerically \citep[e.g.][]{lau07,bar10,zhe10c}.
As argued above, part of
the cause of the line suppression (mostly in the wings) is due to the IGM up
to $\sim$1 Mpc of the source. At this distance, the surface brightness is much
lower than observational thresholds, but could be detected by stacking images
of LAEs \citep{zhe10c}

As seen in e.g.~\Fig{Fscat}, a large scatter between individual sightlines
exists, reflecting the generally quite inhomogeneous IGM. Consequently, one
cannot simply use the calculated transmission function for deconvolving
observed Ly$\alpha$ lines to obtain the intrinsic line profiles, other than in
a statistical sense. With a large sample, however, more accurate statistics on
Ly$\alpha$ profiles could be obtained.
Calculating the transmission functions as an average of all directions, as we
have done in this work, assumes that observed galaxies are randomly oriented
in space, i.e. that there is no selection effects making more or less luminous
directions pointing toward the observer. For LAEs clustered in, e.g., filaments,
the effect of the retarded Hubble flow may be enhanced perpendicular to the
filament, making the galaxies more luminous if observed along a filament than
perpendicular to it \citep[see Figure 9 in][]{zhe10b}.


\subsection{Galactic outflows}
\label{sec:outflow}

As discussed in the introduction, at high redshifts many galaxies are still in
the process of forming and are expected to be accreting gas. In principle, this
should result in a blueshifted Ly$\alpha$ profile, but this is rarely seen.
Evidently, IGM absorption is unable to always be the cause of this missing blue
peak.
On larger scales, mass is observed to be conveyed through large streams of gas;
the cosmic filaments. Although this has been seen only tentatively on galactic
scales, galaxy formation may be expected to occur in a similar fashion.
Indeed, numerical simulations confirm this scenario
\citep[e.g.][]{dek09,goe10}.
Very recently, \citet{cre10} reported on an ``inverted'' metallicity gradient
in three $z\sim3$ galaxies, which they interpret as being due to the central
gas having been diluted by the accretion of primordial gas.
In contrast to gas accreting through a few narrow streams, outflows are more or
less isotropic.
Even bipolar outflows have a rather large opening angle, and thus
the probability of a sightline towards a galaxy passing through outflowing gas
is larger than passing through infalling gas.
Thus, it may be expected that more observations are obtained of Ly$\alpha$
profiles lacking the blue peak than lacking the red peak.

The fact that starbursts are needed to generate large outflows also imposes a
bias on the observations;
even if outflows happen only during relatively short phases in the early life
of a galaxy (for LBGs, \citet{fer06} found that a typical starburst phase lasts
only about $30\pm5$ Myr), such galaxies are more likely to appear in surveys,
simply because they are more luminous than those with small SFRs.



\subsection{Transmitted fraction of Ly$\alpha$ photons}
\label{sec:fIGM}

Even though absorption in the IGM does not alter the line shape drastically
at an intermediate redshift of 3.5, it reduces the intensity by roughly
one-fourth, as seen from \Tab{fIGM}.
This fraction is not a strong function of the size of a galaxy, but since the
spectra emerging from larger galaxies tend to be broader than those of smaller
galaxies, a comparatively larger part of the small galaxy spectra will fall in
the wavelength region characterized by the dip in the transmission function.
Hence, on average the IGM will transmit a larger fraction of the radiation
escaping larger galaxies. 

One-fourth is not
a lot, but since it is preferentially blue photons which are lost, the
spectrum may become quite skewed when traveling through the IGM.
The lost fraction $f_{\mathrm{IGM}}$ is in addition to what is lost internally
in the galaxies due to the presence of dust.
As mentioned in Section \ref{sec:eff}, dust tends to make the line profile
more narrow. For galaxies with no dust, the lines can be very broad.
In this case, $f_{\mathrm{IGM}}$ will be slightly higher, since for broad
lines a relatively smaller part of the spectrum falls on the dip seen in
$\Flam$.
Note however that in general distinguishing between dust absorption and
IGM effects will be rather tricky, if not impossible
\citep[see also][]{day10a}.

As expected, at higher redshifts the IGM is more opaque to Ly$\alpha$ photon;
at $z = 5.8$ and $z = 6.5$ the intensity was found to be reduced by
approximately 75\% and 80\%, respectively.
At $z = 6.5$ the blue wing is completely lost. In most cases this is true also
at $z = 5.8$.
However, as is seen from the gray area representing the 68\% confidence
interval, in some cases an appreciable fraction of the blue wing can make it
through the IGM.


\subsection{The significance of dust}
\label{sec:dust}

The IGM calculations performed in this work are all neglecting the effect of
dust.
However, dust is mostly (although not completely) confined to the galaxies,
especially the central parts, and since by far the greatest part of the
distance covered by a given sightline is in the hot and tenuous IGM, one may
expect this to be a fair approximation.

This anticipation was confirmed following the dust model of \citet{lau09b}.
In short, dust density scales with gas density and metallicity, but is reduced
in regions where hydrogen is ionized to simulate the effect of dust destruction
processes. The factor $\nhi\sigma(\lambda)$ in \Eq{tau} is then replaced by
$\nhi\sigma(\lambda) + n_{\mathrm{d}}\sigma_{\mathrm{d}}(\lambda)$,
where $n_{\mathrm{d}}$ and $\sigma_{\mathrm{d}}(\lambda)$ are the density and
cross section of dust, respectively.
The resulting decrease in transmission is at the $10^{-4}$ to $10^{-3}$ level.


\subsection{``Early'' vs.~``late'' reionization}
\label{sec:10vs6}

As is evident from \Fig{Songaila}, although none of the models really
\emph{fit}, the early and late reionization bracket the observations.
Model 1, with $\sigma_8 = 0.74$ provides a somewhat better fit to the
observations than Model 2 with $\sigma_8 = 0.9$. Due to the more clumpy
structure of the latter, galaxies tend to form earlier, rendering the IGM more
free of gas and thus resulting in a slightly more transparent Universe.
However, this should not be taken to mean that the lower value of $\sigma_8$
is more realistic, since a higher $\sigma_8$ could be accounted for by a
(slightly) smaller $\zre$.

Investigating the LAF is not the only way of probing the EoR. The WMAP
satellite detects the effects of the Thomson scattering of cosmic microwave
background (CMB) radiation on free electrons. Since the number density $n_e$ of
free electrons increases as the Universe gets reionized, a signature of the EoR
can be obtained by measuring the total optical depth $\tau_e$ to Thomson
scattering. The latest results \citep{jar10} give $\tau_e = 0.088 \pm 0.015$,
implying for their model of the EoR that the Universe was reionized at
$z_{\mathrm{reion}} \sim 10$--11.

The optical depth to Thomson scattering in a short interval $dt$ of time is
\begin{eqnarray}
\label{eq:dtau}
\nonumber
d\tau & = & n_e(z)\, \sigma_{\mathrm{T}}\, c\, dt\\
      & = & n_e(z)\, \sigma_{\mathrm{T}}\, c\, \frac{1}{(1+z) H(z)}\, dz,
\end{eqnarray}
where $\sigma_{\mathrm{T}}$ is the Thomson scattering cross section.
That is, given an electron density
history, the resulting total $\tau_e$ can be calculated by integrating
\Eq{dtau}.

\begin{figure*}
\epsscale{1.0}
\plotone{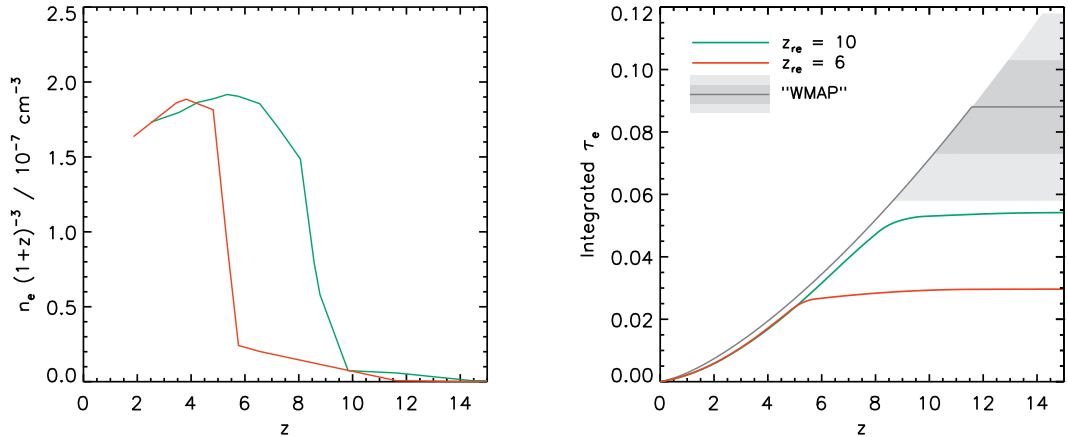}
\caption{\emph{Left:} Volume averaged, comoving electron density $n_e/(1+z)^3$
         as a function of redshift $z$ for Model 2 (\emph{green}) and 3
         (\emph{red}) which have $\zre = 10$ and $\zre = 6$, respectively.
         Both models are characterized by a quite sharp increase in
         $n_e$ shortly after the onset of the UVB. For Model 2
         the EoR is seen to take place around $z \sim 8.5$, while for
         Model the EoR lies at $z \sim 5.5$.
         \ifnum\astroph=1 \\ \fi
         \emph{Right:} Integrated optical depth $\tau_e$ of electrons as a
         function of redshift $z$ for the two models.
         The \emph{dark gray} line with the associated \emph{lighter gray} 68\%
         and 95\%
         confidence intervals indicates the $\tau_e$ history required
         to reach the optical depth measured by WMAP, if an instant
         reionization is assumed.
         As expected, none of the models are able to reach the $0.088\pm0.015$
         inferred from the WMAP results, although the $\zre = 10$ model is
         ``only'' $\sim$2$\sigma$ away.
         Since the $n_e$ data do not extend all the way to $z = 0$, a fiducial
         value of $n_e(z=0) = 1.5\times10^{-7}$ cm$^{-1}$ has been used. The
         exact value is not very imporant, since the proper density at low
         redshift is very small.
         At low redshift, the model curves lie slightly below
         the theoretical curve. This is due to a combination of the models
         including helium
         ionization, releasing more electrons, and star formation and gas
         cool-out, removing free electrons.}
\label{fig:WMAP}
\end{figure*}

\Fig{WMAP} displays the average comoving number density $n_e$ of electrons as a
function of redshift, as well as the corresponding total optical depth
$\tau_e$ of electrons. The results for Model 2 and Model 3 are shown; those for
Model 1 lie very close to those of Model 2.
Going from larger toward smaller redshifts, both models are seen to be
characterized by a roughly constant and very low density of electrons before
reionization, then a rapid increase not \emph{at}, but shortly \emph{after}
$\zre$, and finally a slow decrease,
due to subsequent gas cool-out and resulting galaxy formation.
The rapid increase marking the EoR lasts approximately 100 Myr.
The difference in ionization history for the pseudo- and the realistic UV RT
schemes is not critical, although in the latter the creation of ionized bubbles
around stellar sources causes a slightly earlier EoR than the pseudo-RT is able
to, especially in Model 3 ($\zre = 6$).

Also shown in \Fig{WMAP} is the corresponding $\tau_e$ history that would
prevail in the hypothetical case of an \emph{instant} reionization, where the
Universe is fully neutral before, and fully ionized after, some redshift
$z_{\mathrm{reion}}$ fixed to make the total $\tau_e$ match the value measured
by WMAP.
In this model, no gas is assumed to be locked up in stars, and $n_e$ is thus
given by
\begin{equation}
\label{eq:ne}
n_e(z) =  \left\{ \begin{array}{ll} 0 & \textrm{for } z > z_{\mathrm{reion}}\\
          \frac{\psi \Omega_b \rho_c (1+z)^3}{m_{\mathrm{H}}}
                                      & \textrm{for } z \le z_{\mathrm{reion}},
          \end{array} 
          \right.
\end{equation}
where $\psi = 0.76$ is the mass fraction of hydrogen, $\Omega_b = 0.046$ is the
baryonic energy density parameter \citep{jar10}, $\rho_c$ is the critical
density of the Universe, and $m_{\mathrm{H}}$ is the mass of
the hydrogen atom.

As seen from the figure, not even the early reionization model is
able to reproduce the optical depth probed by WMAP.
However, for a model with an even earlier EoR, the transmission $\T$ of the
IGM would be even farther off the observational data, as seen from
\Fig{Songaila}.
In general, there seems to be a significant disagreement between the WMAP
results and the QSO results concerning the redshift of the EoR.
Many authors have tried to resolve this apparent discrepancy,
e.g.~\citet{wyi03} and \citet{cen03} who considered a \emph{double}
reionization, first at $z \sim 15$--16 and later at $z \sim 6$.

Notice nevertheless that the total $\tau_e$ of the $\zre = 10$ model is not
entirely inconsistent with the WMAP-inferred value, being roughly 2$\sigma$
away.
Furthermore, since only
the \emph{total} optical depth is measured by WMAP, the exact history of
the EoR is obviously less certain. Generally, either an instant reionization
must be assumed, or perhaps a two-step function to make the EoR slightly
extended, and possibly with an additionally
step at $z \sim 3$ to account for helium reionization.

The above results are at odds with the interpretation of the CMB EE polarization
maps as showing that the reionization of the Universe was complete at
$z = 10.5$ \citep{jar10}, regardless of which model is assumed for the EoR.
However, many factors enter the conversion of the
polarization maps into an optical depth, and
the model of this EoR may be too simplistic.
Compared to WMAP, the recently launched
Planck\footnote{http://www.rssd.esa.int/Planck/}
satellite has a much higher resolution and
sensitivity; when Planck data become available, these issues may be solved as
much tighter constraints can be put on parameters like $\tau_e$ \citep{gal10}.




\section{Summary}
\label{sec:sum}

Simulating sightlines through cosmological hydrosimulations,
we have investigated the transmission through the IGM of light in the vicinity
of the Ly$\alpha$ line. The high resolution of the cosmological simulations
combined with the adaptive gridding for the RT allows us to probe the
velcity field around the galaxies in great detail.
While earlier studies \citep{zhe10a} have shown that this approach of simply
multiplying an $e^{-\tau}$ factor on the intrinsically emitted Ly$\alpha$ line
is a poor approximation for the observed line profile and transmitted fraction,
we have argued that when the circumgalactic environs are sufficiently resolved
and combined with realistically calculated intrinsic Ly$\alpha$ lines,
this approach should be valid.
Special emphasis was put on how Ly$\alpha$ line profiles
emerging from high-redshift galaxies are reshaped by the surrounding IGM.
In general a larger fraction of the blue side of the line center $\lambda_0$
is lost as one moves toward higher redshifts.
At $z \gtrsim 5$, almost all of the light blueward of $\lambda_0$ is lost,
scattered out of the line of sight by the high neutral fraction of hydrogen.
However, even at relatively low redshift more absorption takes
place just blueward $\lambda_0$.
A transmission function $\Flam$ was calculated, giving the fraction of light
that is transmitted trough the IGM at various epochs. At all redshifts where
some of the light blueward of $\lambda_0$ \emph{is} transmitted (i.e. at
redshifts below $\sim$5.5), a significant dip with a width of the order 1 {\AA}
is seen.
The origin of this dip is a combination of an increased density of neutral
hydrogen and a retarded Hubble flow in the vicinity of the galaxies.

This extra absorption may in some cases be the reason that the blue peak of an
otherwise double-peaked Ly$\alpha$ profile is severely reduced, or lacking.
Nevertheless, it is not sufficient to be the full explanation of the numerous
observations of $z \sim 2$--4 Ly$\alpha$ profiles showing only the red peak.
The outflow scenario still seems a credible interpretation.

Combining the inferred transmission functions with simulated line profiles,
the fraction of Ly$\alpha$ photons that are transmitted through the IGM at
$z \sim 3.5$, 5.8, and 6.5 was computed, and was found to be
$f_{\mathrm{IGM}}(z=3.5) = 0.77_{-0.34}^{+0.17}$,
$f_{\mathrm{IGM}}(z=5.8) = 0.26_{-0.18}^{+0.13}$, and
$f_{\mathrm{IGM}}(z=6.5) = 0.20_{-0.18}^{+0.12}$, respectively.
This is in addition to what is lost internally in the galaxies due to dust.
The standard deviations were found to be dominated by sightline-to-sightline
variations rather than galaxy-to-galaxy variations.

Considering the average fraction of light far from the line on the blue side
transmitted through the IGM, and comparing to the comprehensive set of
observations of the LAF by \citet{son04}, we constrain the EoR to have
initiated between $\zre = 10$ and $\zre = 6$, corresponding the having
ionized a significant fraction of the Universe around $z \sim 5.5$ and
$z \sim 8.5$, respectively.
Even though the ``early'' models of $\zre = 10$ produces a slightly too
transparent Universe when comparing to the LAF,
the optical depth of electrons is too low
when comparing to the observations of the CMB by the WMAP satellite, possibly
indicating a too simplistic interpretation of the CMB polarization.

\acknowledgments
We are grateful to Antoinette Songaila Cowie for letting us reproduce her
observational data, to Andrea Ferrara for helpful comments on the electron
optical depth,
to Zheng Zheng for pointing out errors and deficiencies in
the first draft, and to the anonymous referee for comments leading to a
substantially extended and better paper.

The simulations were performed on the facilities provided by the Danish Center
for Scientific Computing.

The Dark Cosmology Centre is funded by the Danish National Research Foundation.

PL acknowledges fundings from the Villum Foundation.
%



\end{document}